To the University of Wyoming:

The members of the Committee approve the thesis of Anh Nguyen presented on May 6$^{th}$, 2014.

Dr. Amy Banic, Chairperson

Dr. Steven Barrett, External Department Member

Dr. Jeff Clune

APPROVED:

Dr. James Caldwell, Department Head, Department of Computer Science

Dr. Khaled A.M. Gasem, College Dean, College of Engineering

Nguyen, Anh, <u>3DTouch: Towards a Wearable 3D Input Device for 3D Applications</u>, M.S., Department of Computer Science, August, 2014.

Three-dimensional (3D) applications have come to every corner of life. We present 3DTouch, a novel 3D wearable input device worn on the fingertip for interacting with 3D applications. 3DTouch is self-contained, and designed to universally work on various 3D platforms. The device employs touch input for the benefits of passive haptic feedback, and movement stability. Moreover, with touch interaction, 3DTouch is conceptually less fatiguing to use over many hours than 3D spatial input devices such as Kinect.

Our approach relies on relative positioning technique using an optical laser sensor and a 9-DOF inertial measurement unit. We implemented a set of 3D interaction techniques including selection, translation, and rotation using 3DTouch. An evaluation also demonstrates the device's tracking accuracy of 1.10 mm and 2.33 degrees for subtle touch interaction in 3D space.

With 3DTouch project, we would like to provide an input device that reduces the gap between 3D applications and users.

# 3DTOUCH: TOWARDS A WEARABLE 3D INPUT DEVICE

# FOR

# 3D APPLICATIONS

by

Anh Mai Nguyen

A thesis submitted to the Department of Computer Science

and the University of Wyoming

in partial fulfillment of the requirements

for the degree of

MASTER OF SCIENCE

in

COMPUTER SCIENCE

Laramie, Wyoming

August 2014

COPYRIGHT PAGE

# Acknowledgements

I would like to express my deepest appreciation and gratitude to my committee chairperson, Professor Amy Banic, for her continuous support and advice through the learning process of this master thesis. Without her guidance and persistent help, this thesis would not have been possible.

I would like to thank the rest of my thesis committee: Professor Steven Barrett and Professor Jeff Clune for their encouragement, constructive comments and suggestions, and especially precious advice.

My sincere thanks also go to Professor Jerry Hamann, and Mr. George Janack at the Electrical and Computer Engineering department, and Mr. John Kicklighter for their kind and valuable advice and suggestions regarding electrical circuitry.

I would like to thank the UW School of Energy for partly funding my work through the Research Assistantship granted to me.

Last but not least, I owe more than thanks to my beautiful wife, Duong Do, and my family for their love, understanding, and support for me in completing this thesis.

# Table of Contents

# 1  Introduction

Virtual Reality (VR) is a computer-generated simulation of a 3D environment in which users can perceive and react as if they are in a real environment. A virtual environment (VE) is primarily experienced through the two senses of sight and sound. A good VR system provides users with high sense of presence so that they suspend their beliefs and accept it as a real environment. Hence, VR is a useful platform for gaming, training, education, visualization and many other applications.

VR has gained more and more interest in the last decade in many different disciplines. VR has been considered an effective research tool for kinesiology therapists [1]; and the future of clinical psychology [2]. Its usage also expands to scientific visualization [3], especially with the emergence of Cave Automatic Virtual Environment (CAVE) [4] which allows users to immerse themselves in a virtual world in a physical room with walls projected with computer-generated imagery. Besides becoming a popular facility at academic research labs, VR is also widely used in the engineering industry for design and manufacturing applications at BMW, Volkswagen and many others [5].

There are a variety of different 3D input devices that have been used to interact with immersive virtual environments. They range from 2D to 3D, custom-made by computer scientists to popular commercial devices such as Wiimote and Kinect [6]. Two current open questions [7] in the field of 3D User Interfaces (3DUI) are:

1) *How should we design 3D input devices?*
2) *What are the most appropriate mappings between 3D input devices, displays, and interaction techniques?*

As of the time being, there is no standard set of input devices for virtual environments. The choice depends on the application itself, user preferences, and the device availability. Recent years have witnessed a wide variety of input devices to interact with 3D applications [8]. Desktop input devices such as traditional mice, keyboards, or 3D mice (e.g., 3Dconnexion SpaceNavigator) provide stability and accuracy; however, they are not portable for spatial environments such as the CAVE. Mobile touch devices provide intuitive and direct input [9], but the working space is limited within the screen area. While voice input is convenient, it is not intuitive for users to give voice commands for performing complex 3D interaction tasks (e.g., rotate the red cube 60 degree around z-axis). Although these devices have their unique advantages, they are usually designed for a single certain platform. There is a need for a universal 3D input device that works across multiple 3D platforms in order to bring users closer to 3D applications.

One input method to interact with VE is using 3D mid-air gestures, which are popularized by commodity devices like Kinect and Wiimote. Although these devices are relatively intuitive, natural and easy to use [10], they have a disadvantage that mid-air gestures can be quite tiring when performing for long hours as the interactions require users to stretch their arms out away from the body. In 2011, a study showed that 3D mid-air gestures with bare hands are more tiring than 1D and 2D gestures with hand-held input devices (e.g., smartphones, or remote controls) [11]. Commodity devices were originally designed for gaming and exercise purposes, and were later adopted to VR. Hence, they are not intended for long-hour use for purposes beyond entertainment.

Touch interaction is another way to interact with 3D applications. Unlike spatial interaction, touch interaction has a subtle neat advantage that users can feel natural passive haptic feedback on the skin via the sense of touch. Touch gestures are conceptually less fatiguing than 3D mid-air gestures. Moreover, the touch surface keeps the hand steady and thus increases the stability and accuracy of finger movements. A variety of creative research works have then brought touch interaction to surfaces that are not inherently touch-sensing capable such as tables [12], walls [13], clothes [14], skin [15], [16], [17], conductive surfaces [17] (e.g., the metal door knob, and even liquids), or virtually any flat surface using a combination of a depth-sensing camera and a projector [15], [18].

We were motivated to build a novel 3D input device that can be used universally across multiple platforms, and for many hours with the least possible fatigue. We present 3DTouch, a thimble-like 3DTouch input device worn on the user's fingertip. 3DTouch is self-contained, and universally works on various platforms (e.g., desktop, and CAVE). The device employs touch input for the benefits of passive haptic feedback, and movement stability. On the other hand, with touch interaction, 3DTouch is conceptually less fatiguing to use over many hours than spatial input devices.

Another advantage of 3DTouch over some of the existing 3D input devices is that it is not subject to the occlusion problem. A common setup of an immersive virtual environment includes an optical tracking system, which has an inherent problem of occlusion described in section 2.1.3. The cameras of the tracking system need to be able to see an object in order to track its position. 3DTouch is not subject to such a problem as a relative position can be derived by fusing data from its own sensors.

3DTouch allows users to perform touch interaction on many surfaces that can be found in an office environment (e.g., mousepad, jeans, wooden desk or paper). When mounted on the tip of index finger, the user can perform touch interaction on the other hand's palm, which serves as the touch pad. 3DTouch fuses data reported from a low-resolution, high-speed laser optical sensor, and a 9-DOF inertial measurement unit (IMU) to derive relative position of a pointer in 3D space. The optical sensor, usually found in traditional computer mice, determines the direction and magnitude of movement of the pointer on a virtual 2D plane. And the 9-DOF IMU determines the orientation of the plane. Since we would like to keep the 3DTouch interface simple with no buttons, a gesture recognition engine was developed to allow users to make gestural commands. Based on the data from the optical sensor, we used classification techniques to reliably recognize simple gestures such as: tap, double-tap, and press gesture.

This thesis describes the work previously reported in the following two publications: [19] and [20].

# 2 Background

3DTouch is an interdisciplinary research project that crosses various fields. In this section, we review the related literature in the areas of Immersive Virtual Environments (IVE), 3D Input Devices, and Wearable Computing.

## 2.1 Immersive Virtual Environments

"Virtual Reality", synonymous with "Virtual Environment", usually used to describe a synthetic, spatial and 3-dimensional world seen from a first-person point of view.

### 2.1.1 Concept and Applications

The view in a virtual environment is under the real-time control of the user [6]. There are two usual implementations of an IVE. The first of these involves placing multiple projection screens and loudspeakers around the user. A now popular design in many institutions and organizations is the CAVE [4] which is a moderately sized cubical room with walls, floor and ceiling back-projected with computer-generated imagery. The system is usually integrated with a tracking system that measures the changing position and orientation of the user's head within the room to generate imagery in proper viewing perspective. Users in the room wearing 3D stereoscopic glasses can then perceive the virtual world around in 3D. There can be more than one user simultaneously using the CAVE; however, only one's position will be used to generate proper viewing perspective.

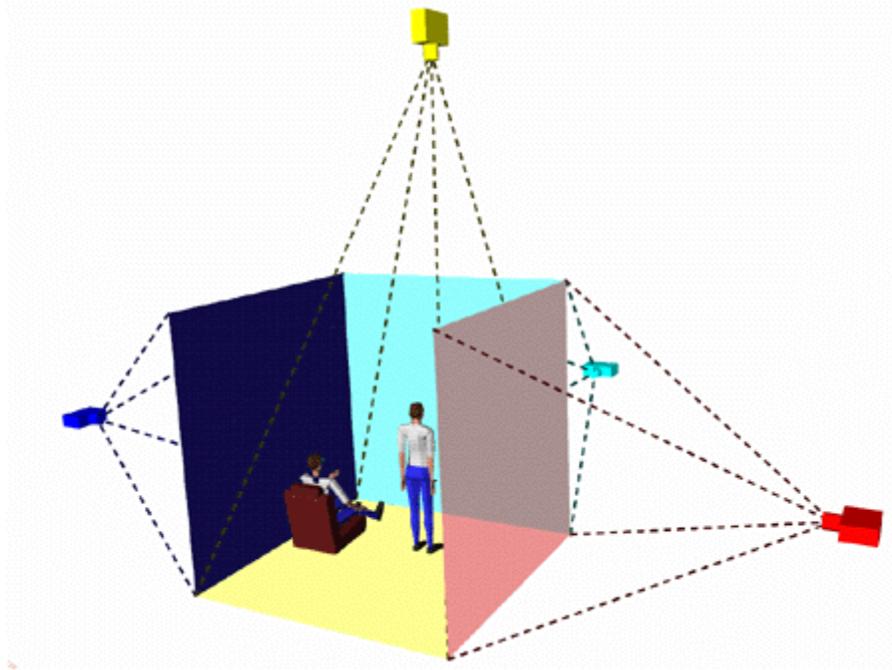

*Figure 1 - Cave Automatic Virtual Environment with projected walls. Image courtesy: Australian National University.*

The second, more common and less costly implementation of IVE consists of the use of a head-mounted display (HMD) originally invented by [21], in conjunction with a head tracker. At any given moment, the computer generates and outputs the visual and auditory imagery to the HMD from a perspective that is based on the position and orientation of the user's head.

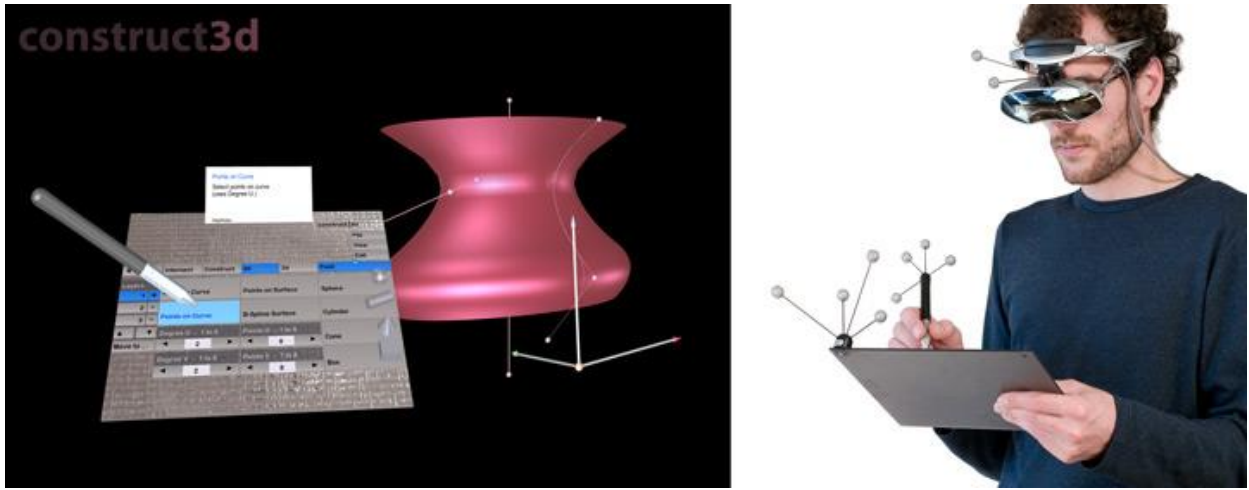

*Figure 2 - A user (right side) wearing an HMD and manipulating a menu widget in the virtual world (left side). The HMD, pen, and tablet are attached with markers so that they can be tracked by an optical tracking system. Image courtesy: IOTracker.*

Since immersive virtual environments can provide users with a high sense of presence, they have been used as a research tool in various fields such as psychology [22], or rehabilitation [1]. Other well-known applications of IVE range from visual simulation to gaming entertainment [23], and also scientific visualization [3]. The purpose of our 3D input device is to provide scientists with better user interfaces to better explore scientific visualization in IVE.

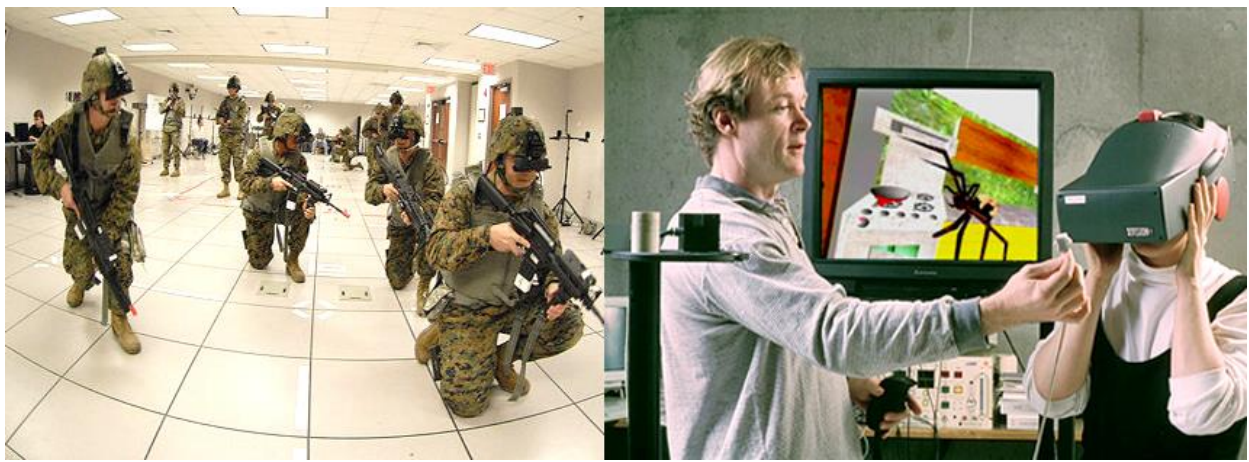

*Figure 3 - Virtual Reality applications in military training (left) and psychological therapy (right).*

***On the left**: a group of solders participating in a military training session wearing HMDs. Image courtesy: Real Vision FZ LLC.*

***On the right**: a patient is receiving treatment to cure her fear of spiders. Image courtesy: HITLab Washington University.*

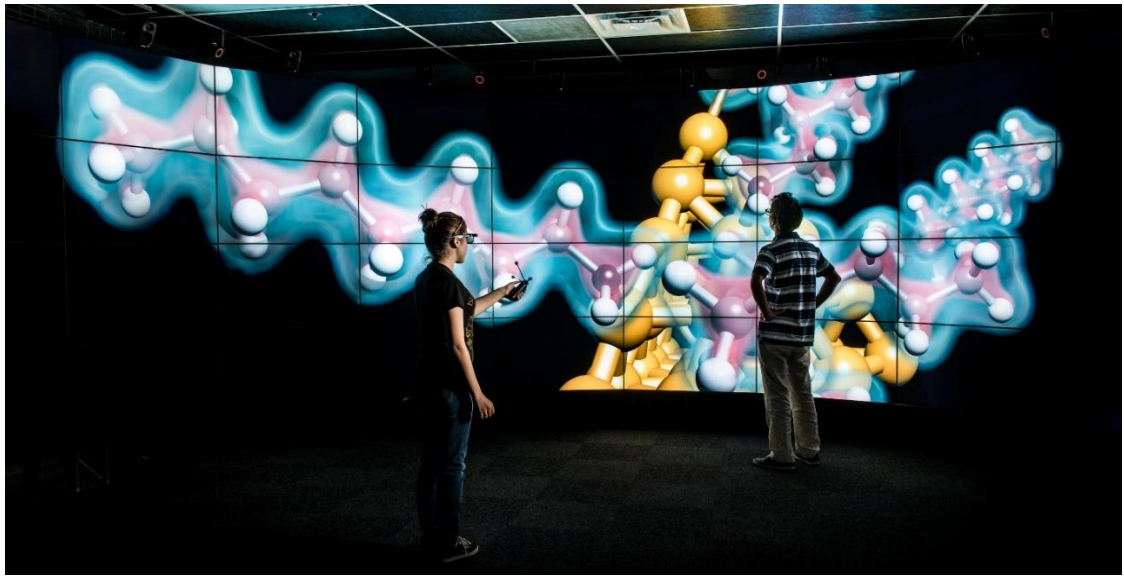

*Figure 4 – Visualization of molecular structures at Electronic Visualization Lab, UIC. Image courtesy: Khairi Reda*

As [3] pointed out, the limitations of scientific visualization using IVE at the time are data management, rapid prototyping, networks, architecture, visual displays and input devices. Input devices, particularly, needed higher accuracy, longer-range tracking capability, algorithms for inferring user intentions, improved gloves, better button devices, speech recognition, and design guidelines. Many of these limitations have been improved over the years; and our proposed solution is another attempt to improve the current input devices for visualization scientists.

### 2.1.2 Existing 3D Input devices for IVE

Selecting appropriate output devices is an important component of designing, developing and using 3D application because they are the primary means of presenting information to the user. Output devices directly impact users' sense of presence and how they perceive information. In immersive visualization, a common setup of output devices would include a HMD or 3D glasses that enable users to perceive the virtual world in three dimensions. However, an equally important part of the application design is choosing the appropriate set of input devices that allow users to communicate effectively and efficiently with the application. There are many different types of input devices to choose from when developing a VR application, and some devices may be more appropriate for certain tasks than others.

According to [6], 3D input devices can be broken down into the following categories:

- Desktop input devices
- Tracking devices
- 3D mice
- Special-purpose input devices
- Direct human input

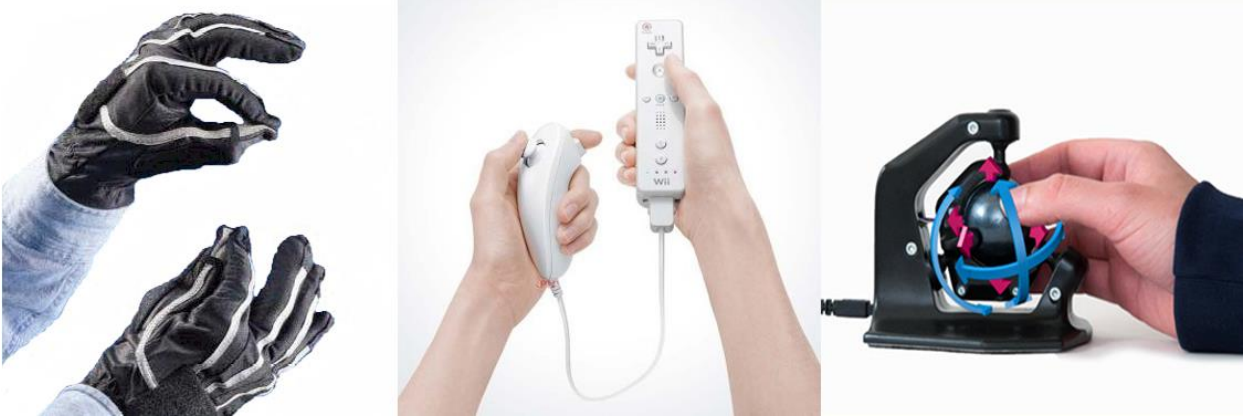

*Figure 5 - Three of many existing popular VR input devices. From left to right:*

1. *Pinch Glove: an input device that determines if a user is touching two or more fingertips together. These gloves carry conductive material at each of the fingertips so that when two fingertips touch each other, an electrical contact is made. Image courtesy: FakeSpace Labs.*

2. *Wii Remote: is the primary controller for Nintendo Wii console. Its main feature is motion sensing capability which allows users to interact with and manipulate items on screen via gesture recognition with the controller through the use of accelerometer and optical sensors. Image courtesy: Nintendo.*

3. *Axsotic 3D mouse: tracks a 40mm ball, which can be twisted, pushed, pulled, lifted and so on inside a sensor-laden cage. Optical tracking watches for rotation in three axes, while magnets track zoom and pan in three axes. This device allows desktop users to manipulate objects in 3D in a stable form. Image courtesy: Axsotic.*

In the next sections, we will introduce several existing input devices commonly used nowadays to serve as the base for the motivation of our proposed solution.

### 2.1.3 Optical Tracking System

One of the important aspects of 3D interaction in virtual worlds is providing a correspondence between the physical and virtual environments. Hence, typically in a CAVE, the system would consist of an optical motion tracking system [4] which tracks the position and orientation of the user's head in order to generate correct imagery output to the projected walls. While vision-based tracking systems have advantages that:

- Users can be completely untethered from the computer (unlike mechanical tracking)
- Relatively high sampling rates compared to acoustic tracking
- Not affected by external noise and acoustically reflective surfaces (like acoustic tracking) and ferromagnetic objects in the room (like magnetic tracking)
- Less error (sensor biases, noise, and drift) than inertial tracking

Below is the image from OptiTrack demonstrating a typical setup of the optical tracking area in our 3DIA lab at the University of Wyoming, and this is also the setup we use for our project:

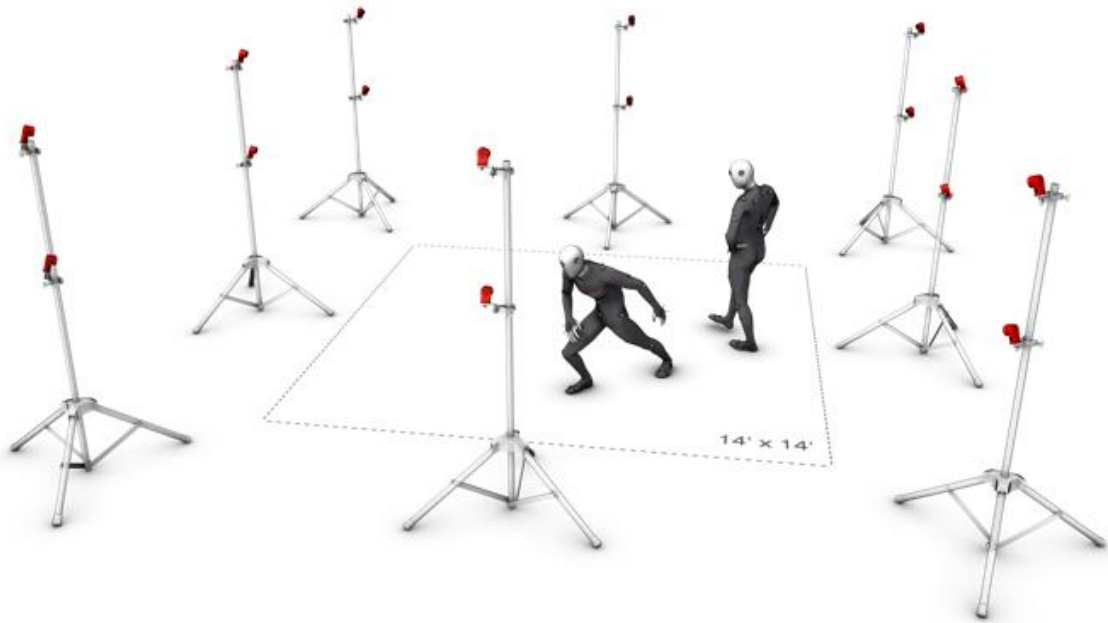

*Figure 6 – A typical setup of an Optical Tracking system which consists of cameras that define a tracking volume. Trackable objects such as Head-mounted displays with markers will be recognizable by the system if moving within the pre-defined area. It is also possible to track the movement of human body parts if they put reflective markers on their body. Image courtesy: NaturalPoint.*

However, the major disadvantage of vision-based tracking system is occlusion [24]. The cameras cannot see a certain part of the user's body that is occluded by other parts. Moreover, in a multi-user environment like CAVE, one user may block others from being tracked by cameras while both interacting with the IVE. Our proposed input device is not based on this optical tracking mechanism, hence not prone to this major drawback.

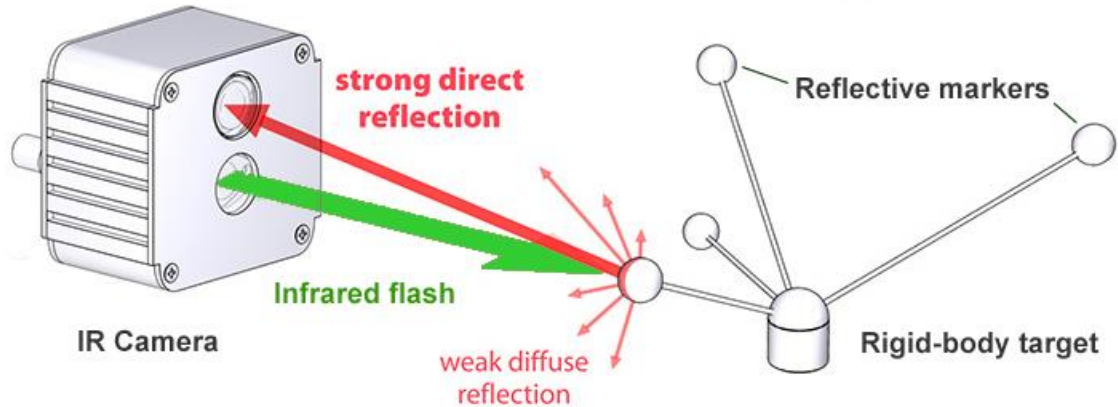

*Figure 7 – Optical tracking mechanism: IR Camera emits an infrared beam to the tracking area. When the beam hits the reflective markers attached to the rigid-body being tracked, a strong direct reflection ray will be reflected back to the camera. The camera then can determine the marker's position in the space. The configuration of the reflective markers helps cameras distinguish between different registered rigid bodies. An object cannot be tracked if the line of sight between the markers and the camera is blocked. Image courtesy: IOTracker.*

## 2.2 Three-dimension Input Devices

### 2.2.1 3D wands

Current technologies have seen the popular usage of 3D input devices like the Nintendo Wiimote as a tool for 3D interaction in IVE [25]. Wiimote is one of the most popular devices which have been used for a variety of purposes, such as for gesture recognition-based applications [26], [27]. While a variety of motion sensing devices have emerged over the years, the advantage of the Wiimote is that it is a low-cost wireless device that features an infrared sensor with accelerometers, vibration feedback, speaker, and easy-to-use buttons all within a single device. These features could make it more favorable than traditional game controllers. Although Wiimote has its own optical tracking system; however, in a large setup such as in a CAVE, this could be replaced by the CAVE's optical tracking system which is usually more robust with more cameras. Although, the wand has proved to be more intuitive than pointing devices in IVE than wired gloves and 3D mice, but it is not the one causing least fatigue [28]. Our proposed solution enables users to interact with IVE using micro-interactions. These micro-interactions are supposed to cause less fatigue than wrist twisting, or arm-stretching required when interacting with the wand. Also according to the study by [28], participants were observed to turn and rotate the wand within the palm of their hand rather than performing the more natural but equivalent rotation with their wrist or forearm. This counter-intuitive interaction, seems to enable a better performance index for the wand. We rely on this fact to hypothesize that our micro-interactions would yield better performance and cause less fatigue than the wand.

### 2.2.2 Natural User Interface using Gesture recognition

Another currently popular type of motion tracking systems is a line of gesture recognition systems like Kinect, LEAP, PlayStation Move and such with Kinect being a leading system that consists of a depth-camera that enables users to interact with their computer through physical motion without the need for a controller. Kinect technology can interpret specific gestures, making completely hands-free control of electronic devices possible by using an infrared signal projected and camera and a special microchip to track the movement of objects in three dimensions.

A sample image from the database of Kinect gestures [29] shows how Kinect recognizes body gestures:

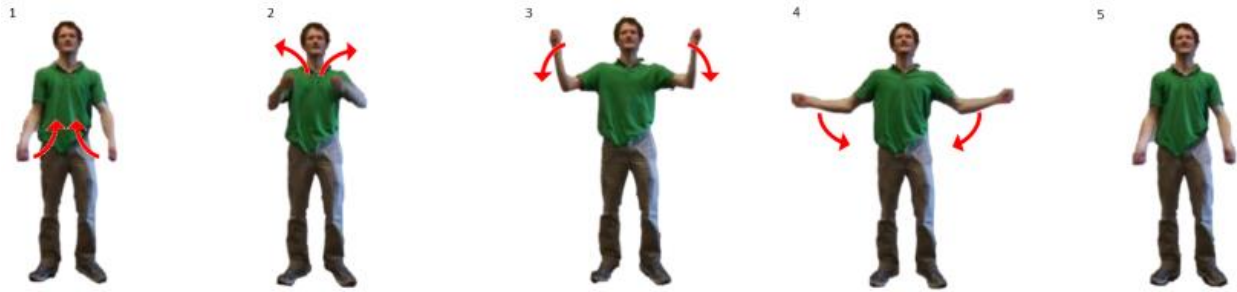

*Figure 8 - Example of gestures recognized by Kinect. Image courtesy: [29]*

So what is the major drawback of Kinect? As demonstrated in the figure above, Kinect focuses on body gestures which usually require users to stand and perform hand movements in the air. While these body movements may not be a big problem for interactive gaming or workout purposes, they could potentially cause fatigue for scientists who use Kinect to work in IVE for long hours. Several studies investigated how Kinect performs against Wiimote in virtual environments, and one of the results found that Kinect is easier to learn and more natural than Wiimote; however, it is not appropriate for a prolonged usage because of the physical effort required to users [30]. A study in 2011 also found that 3D mid-air gestures are more tiring than 2D gestures using hand-held devices such as remote controls, or mobile devices [11].

### 2.2.3 Designing 3D User Interfaces

In IVE, in order to be able to manipulate 3D objects, one generally needs at least six degrees of freedom (6 DOF), three for X, Y, and Z translation, and another three for 3D rotation (pitch, roll, yaw). There is not yet a standard guideline for how the 6 DOF map to the user's body parts. It varies among devices and applications. For example, if a task requires movement in all three dimensions, the input device should support these translations along the three axes simultaneously. If the task requires only two dimensions, as with viewpoint orientation in space, the input device's operational axes should be constrained to prevent unintentional actions [7]. Certainly, designing an input device to handle all 3D tasks would be uneconomical. In this project, we design our input device for the basic tasks of 3D rotation, and translation in visualization applications because such applications are usually used by researchers in various fields and fatigue is a major concern when working long hours.

There are many choices in designing a 6 DOF input devices. The choice on every design dimension may have implications on users' performance. Beside application specific requirements, [31] listed six main usability aspects of a 6 DOF input device:

- **Speed**: Speed of movements performed by the device. Most input devices nowadays are flexible with high speed.
- **Accuracy**: The accuracy of the output of the device. This could be measured with speed.
- **Easy of learning**: Devices like Kinect are intuitive and relatively easy to learn.
- **Fatigue**: This is an important factor to consider and it is related to specific hardware design and interaction techniques developed upon it.
- **Coordination**: is unique to multiple degrees of freedom input control, in this case 6-DOF.
- **Device persistence and acquisition**: is the ease of device acquisition. This is often an overlooked aspect of input device usability.

In designing our proposed input device, we primarily focus on Speed, Accuracy, and especially Fatigue because fatigue is a major concern when working long hours.

## 2.3 Wearable input devices and Sensors

### 2.3.1 Wearable Input Devices

Touch technology has recently been widely available to regular electronics consumers in tablet PCs, smartphones, and laptops. While this technology has succeeded in bringing a new type of input to applications and devices, its utility has a fundamental limitation that the input area is restricted to the touch area.

A variety of technologies have been creatively proposed to bring touch interaction beyond regular surfaces. Cohn et al. [32] explored a new interaction modality that utilizes the human body as a receiving antenna for electromagnetic noise in an environment like a room with walls. By examining the noise picked up by the body, the system can determine whether a user is touching the wall and the position of the contact. Another touch input solution uses an acoustic-based approach [33] by measuring the sound made from the touch contact between fingernails and a variety of surfaces. Skinput [15], similarly, uses a bio-acoustic sensor built into a wearable armband to detect and localize finger taps on the forearm and hand. Acoustic-based implementations, however, can be affected with noise from environment. PocketTouch [14], on the other hand, investigated the capability of an eyes-free touch input method through fabric enclosing the device (e.g. a pocket or bag). While these techniques open up new and interesting interaction opportunities, they cannot sense detailed finger movements required to emulate high-precision touch input.

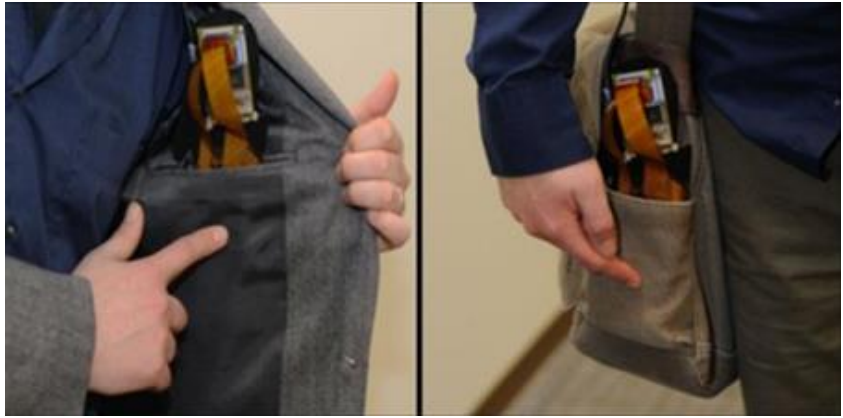

*Figure 9 - PocketTouch enables touch interaction through fabric surfaces. Image courtesy: [14]*

Early work in instrumenting the human finger was conducted with Ring Mouse [6], is a small, ring-like device, with two buttons, worn along the index finger. It uses ultrasonic tracking, but generates only position information. With a similar design to that of Ring Mouse, FingerSleeve uses a 6-DOF magnetic tracker to report position and orientation [34]. The drawback of these devices is that they are not self-contained, and rely on an external tracking system.

Several other wireless finger-worn implementations utilize electromagnetic sensors to enable wireless, unpowered, high fidelity finger input such as Abracadabra [35] and FingerPad [36]. Abracadabra is a magnetically driven input technique for very small mobile devices. The use of a magnet sensed at a distance enables wireless and unpowered input. The interaction of twisting the hand and curling the finger cause the magnetic field to invert enabling different contextual input. FingerPad, on the other hand, uses a Hall sensor on the nail of index finger, and a magnet on the thumbnail. This setup allows pinch gestures between thumb and the index finger. FingerPad achieves 93% accuracy in seated conditions, and 92% in walking conditions. Micro-interactions such as the pinch gestures enabled by FingerPad reduce fatigue significantly compared to interactions that involve the forearm and shoulder. These two approaches while achieving relatively high accuracy are still subject to the inherent problem of being interfered with nearby ferromagnetic or conductive objects. On the other hand, none of the goals of these projects was to build a 3D input device.

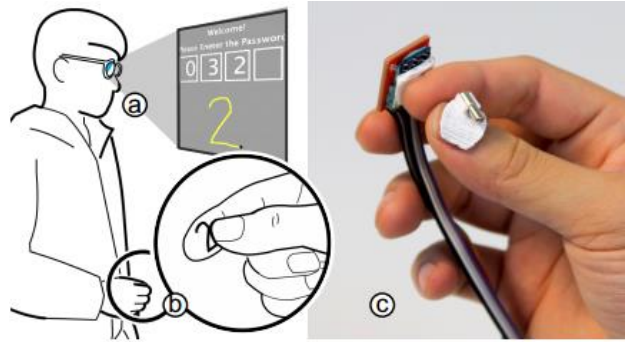

*Figure 10 - FingerPad setup with a Hall sensor grid mounted on the index fingernail and a magnet mounted on the thumbnail. This enables micro-interactions such as pinch gestures. Image courtesy: FingerPad.*

uTrack [37] turns the fingers and thumb into a 3D input device. As a magnet is worn on the thumb, and two magnetometers are worn on the fingers, uTrack is a self-contained 3D input device. However, it is only a 3D pointing device, not a full 6-DOF input device. In addition, the movements of the magnet of uTrack are constrained within a small volume around the magnetometers, and are susceptible to interference within a noisy environment.

Adopting a vision-based motion recognition concept, many researchers have explored various solutions such as [16], or [18] which allow users to perform touch interaction on projected surfaces. The surface is projected with information from a mini projector, and the touch gestures performing on it are recorded by a depth camera like Kinect. Many others have explored the possibility of mounting cameras on the body to enable a variety of different touch regions such as the palm [38], [16] or wrist [39]. The Imaginary Phone [38] uses a 3D camera to detect taps and swipes on a user's palm. OmniTouch [16] uses a depth camera and a micro projector to turn a user's palm into a touch-sensitive display. SixthSense [18] also uses a small projector and a camera worn in a pendant device, to enable touch interactions on and around the body. While such techniques enable convenient and always-available input, they are restricted to the viewing angle of the mounted camera, and can suffer from occlusion.

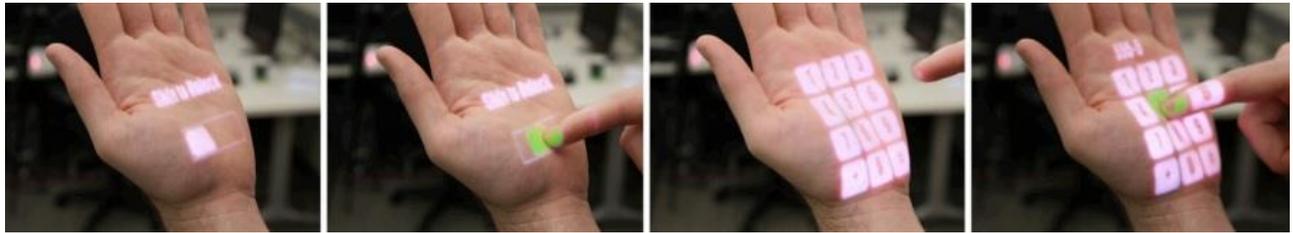

*Figure 11 - OmniTouch an example of implementation that uses a project and a depth-camera mounted on body. Image source: OmniTouch.*

While many approaches focused on improving the sensing surface, [40] inverted the relationship between finger and sensing surface with Magic Finger in which they instrument the user's finger itself instead of the surface it is touching. By instrumenting the finger, Magic Finger enables touch interaction with any surface without the need for body-mounted cameras hence it is not subject to occlusion problem. Magic Finger is a thimble-like device worn on the user's finger. It combines two optical sensors: a low resolution but high speed sensor for tracking movement, and a high resolution camera for capturing detail of the texture of touch surface. Magic Finger can recognize 32 different textures with an accuracy of 98.9%, allowing for contextual input. However, Magic Finger's goal is for ubiquitous use of the device. Our solution, 3DTouch, adopts the same concept of finger instrumentation as Magic Finger, however, we propose to utilize different types of sensors for use in 3D applications.

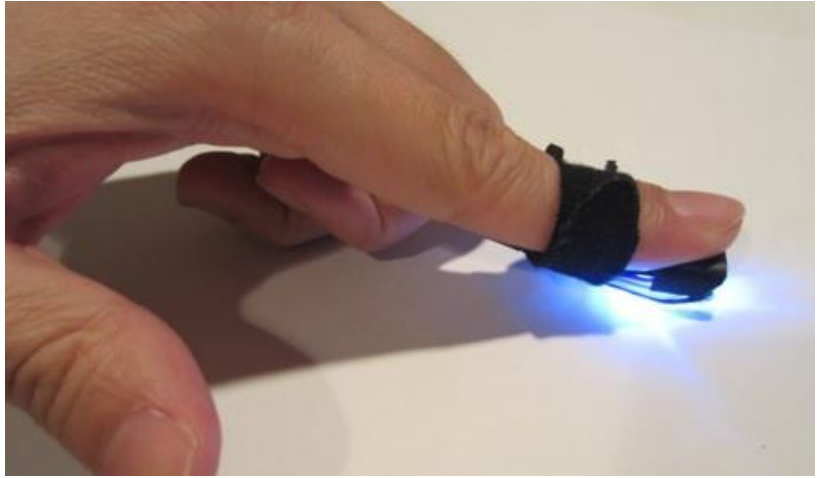

*Figure 12 - Magic Finger with two optical sensors worn on the finger. Image courtesy: Magic Finger.*

### 2.3.2 Optical sensors

Optical sensors residing in mice usually have low-resolution but high speed as their main purpose is to track fast and subtle 2D movements. Earlier optical mice detected movement on pre-printed mouse pad surfaces, whereas modern optical mice work on most opaque surfaces. It is usually unable to detect movement on specular surfaces like glass, although some advanced models can function even on clear glass. Traditional optical sensors are LED-based meaning that the contact area between a sensor and a surface needs to be lit up usually by a Light-Emitting Diode (LED) and photodiodes (similar to MagicFinger). Modern mice nowadays are based on laser diodes which are invisible to human eyes and offer better resolution and precision.

3DTouch uses a laser optical sensor of ADNS-9800 [41] manufactured by Pixart. This sensor uses an infrared laser diode instead of a LED to illuminate the contact surface. The laser illumination enables superior surface tracking compared to LED-illuminated optical mice, while not posing a health risk if pointed to eyes according to its compliance with IEC/EN 60825-1 Eye Safety.

Avago, a leading optical sensor manufacturer, has described in detail the difference between the technologies used in LED-based optical sensor and laser optical sensor [42].

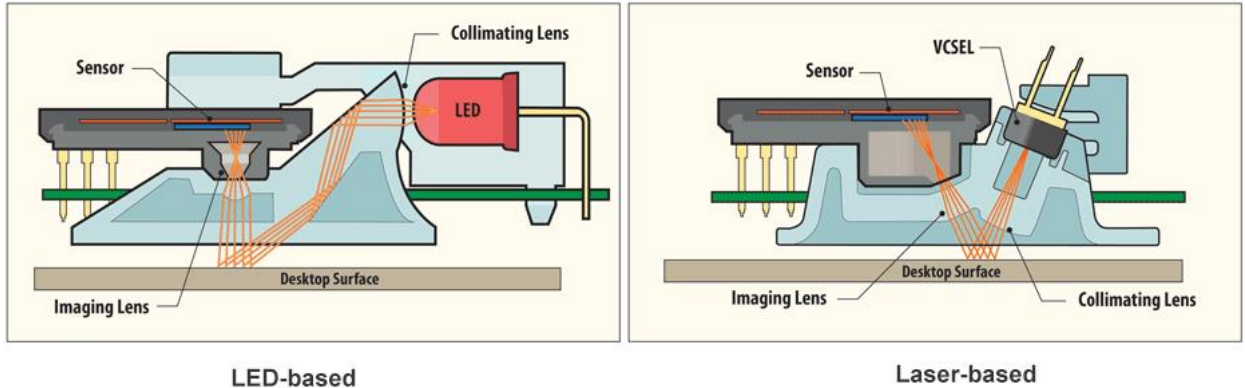

*Figure 13 - LED-based technology (left) uses an LED to illuminate the contact surface, while the Laser-based VCSEL technology uses an invisible laser to illuminate the surface. Image courtesy: [42]*

Their main difference is in the way the surface is illuminated. They both make use of optical lenses to adjust the sensitivity of the mouse relative to the surface.

### 2.3.3   Novel input devices using Optical sensors

Optical sensors are widely used in robotics to measure odometry and velocity of wheeled robots [43]. Similarly, in Ubiquitous Computing, optical sensors are also well explored, aside from MagicFinger, as this type of sensors is usually of small form factor, and easy to acquire. Mouseless [44] is an example that simply mounts the sensor on the side of a laptop to enable mouse interaction without an actual mouse. Soap [45], on the other hand, is a positioning device that creatively places an optical sensor inside a hull made of fabric. As the user applies pressure from the outside, the optical sensor moves independent from the hull. The optical sensor perceives this relative motion and reports it as position input. This original idea; however, only supported pointing interaction in 2D. Another research effort by [39] explored the possibility of placing an optical sensor on the user's arm. Interactions then can be made with the other hand hovering over the sensor.

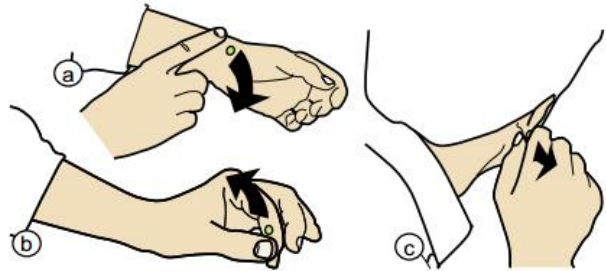

*Figure 14 - The concepts of miniaturizing gestures involve placing tiny optical sensors on the wrist, or finger. Image courtesy: Ni & Baudisch, 2009.*

Although many previous works have investigated the possibility of placing optical sensors on body, they only support 2D gestures, and in our research, we propose to extend this limit to 3D using an additional sensor, a 9-DOF inertial measurement unit.

### 2.3.4 Extra dimensions of Touch Interaction

Many mobile touch devices only utilize the 2D position of a touch contact being made on the surface. However, other auxiliary information of a touch interaction has also proved to be useful such as: the shape [46], [47] or size [48] of the contact region, the orientation of the finger making contact [49], and even the touch pressure [50]. While the size of the contact region was used to improve the precision of selection techniques [48], attributes such as the shape of the contact region [46], [47], orientation of the finger [49], and touch pressure [50] were additional inputs for the application to deliver pseudo-haptic feedback to users.

Using a 9-DOF IMU mounted on the fingernail, 3DTouch leverages the finger orientation to augment the 2D input from the optical sensor into 3D input. The pressure dimension is used to enable *press gesture*, conceptually similar to a mouse-click gesture. Unlike the popular tap gesture on touch devices, press gesture allows the user to make selection commands without lifting the finger off the surface, thus reducing workload for the finger joint.

# 3  3DTouch

This section presents the implementation including the hardware and software parts of our novel 3D input device.

## 3.1  Hardware Prototype

An open problem of spatial tracking is how to build a 6-DOF system that is self-contained, and capable of tracking its own position and orientation with high levels of accuracy and precision [8]. With 3DTouch, our approach is to fuse data from a 9-DOF IMU and a laser optical sensor to derive 3D position and 3D orientation of an object in 3D space.

### 3.1.1  Form Factor

The device needs to be small so that it can fit on the user's finger. We mounted the IMU on top of the fingernail so that we can utilize the finger's orientation. Figure 15 illustrates two ways which the optical sensor can be placed on the fingertip. In form factor 1, if small enough, the optical sensor could potentially be embedded under the fingernail. In form factor 2, the optical sensor could be placed underneath the fingertip pad skin.

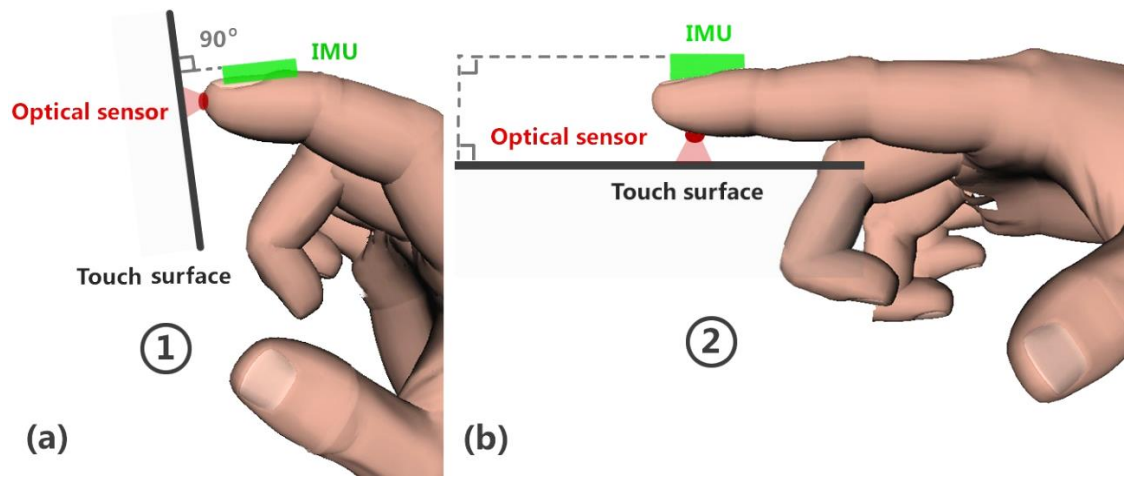

*Figure 15 - Two possible form factors of 3DTouch. (a) Form factor 1: the IMU is placed on top of the fingernail, and the optical sensor is placed on the skin below the fingernail. (b) Form factor 2: The optical sensor is placed underneath the fingertip pad.*

The form factor of the ADNS-9800 optical sensor used does not allow us to implement the prototype as in form factor 1. Our first prototype was implemented according to form factor 2 (Figure 15). The device has the shape of a thimble, it uses an adjustable Velcro strap to hold the sensors. The IMU is mounted on top of the fingernail, and the optical sensor is placed underneath the fingertip pad.

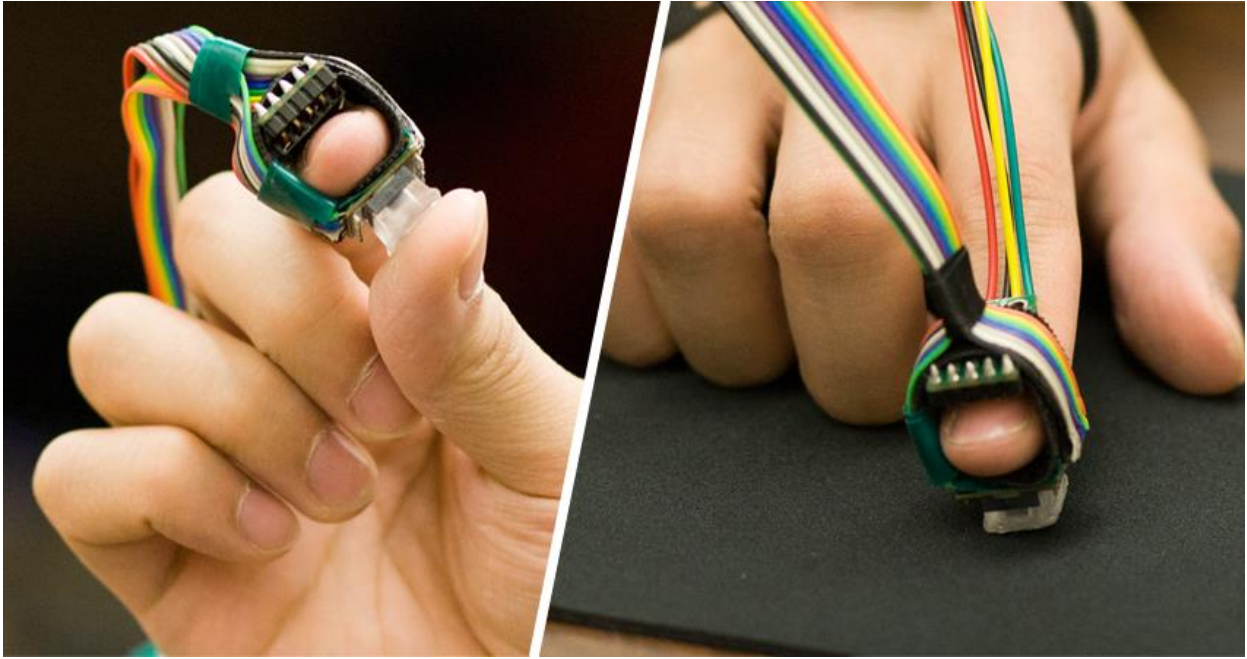

*Figure 16 - Our prototype of 3DTouch following Form factor 2. The device is an adjustable Velcro ring worn on the fingertip.*

The second form factor can be transformed into a third form factor by pushing the device inwards towards the palm (see Figure 17). Wearing 3DTouch as a ring allows users to turn their finger into a pointing device. In this form factor, users do not have to perform touch interaction with a surface. Users can wear 3DTouch as a ring on the index finger or on their thumb as well.

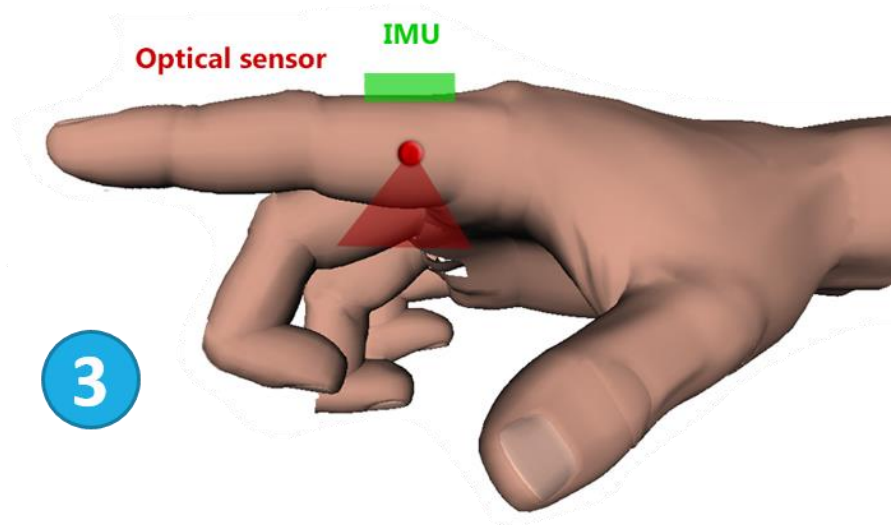

*Figure 17 - The third form factor of 3DTouch. Users can turn their finger into a pointing device, and do not have to perform touch interaction.*

### 3.1.2 Inertial Measurement Unit

An inertial measurement unit, or IMU, is an electronic device that measures and reports on a craft's velocity, orientation, and gravitational forces, using a combination of accelerometers and gyroscopes, sometimes also magnetometers. It is a component that can be found nowadays in devices such as the Wiimote and smartphones.

We used the Pololu MinIMU-9 v2, a 9-DOF IMU, that packs an L3GD20 3-axis gyro, an LSM303DLHC 3-axis accelerometer and 3-axis magnetometer onto a tiny 0.8" x 0.5" board [51]. We chose such an IMU with 9 degrees of freedom because when applying a Kalman filter [52], the estimates of orientation would be more precise than those based on a single measurement alone.

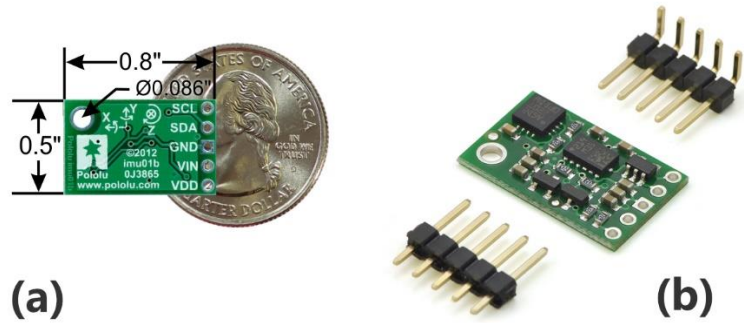

*Figure 18 - The size of the small form factor MinIMU-9 v2 unit. Image courtesy: Pololu Inc.*

*(a) The size of MinIMU-9v2 in comparison with a USD quarter coin. (b) The up side of the unit.*

### 3.1.3 Optical Flow Sensor

We used a Pixart ADNS-9800 laser optical flow sensor with a modified ADNS-6190-002 lens (see Figure 16). This sensor board consists of a camera and also a laser emitter, which emits infrared using laser diode instead of a LED to illuminate the contact surface. The laser illumination complies with IEC/EN 60825-1 Eye Safety, and so does not pose a health risk if pointed to the eyes.

The reason we chose a laser sensor is that they work on a larger number of surfaces than LED-based optical sensors. The ADNS-9800, often found in modern laser gaming mice (e.g., UtechSmart High Precision Laser Gaming Mouse), comprises a sensor and a Vertical-Cavity surface-emitting laser (VCSEL) in a single chip-on-board (COB) package. The sensor is a high resolution (i.e., up to 8200 counts per inch), black-and-white camera (30 x 30 pixels). However, for the purpose of tracking movements, we manually programmed the resolution down to 400 counts per inch (cpi) for a higher frame rate. The details of VCSEL technology are discussed in the section 2.3.2.

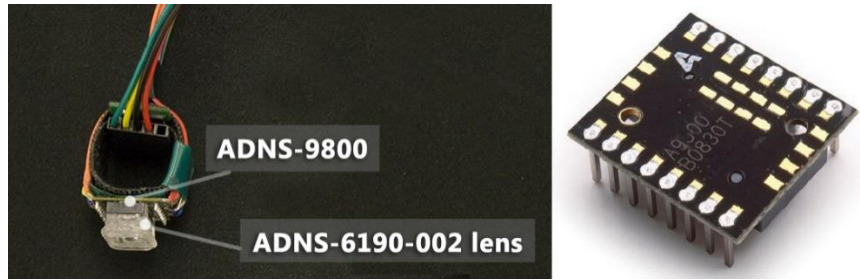

*Figure 19 - The Pixart ADNS 9800 optical sensor and its lens ADNS-6190-002.*

This COB sensor is then wired to an application printed circuit board (PCB) designed according to the schematic diagram in the datasheet [41]. This PCB streams data from the sensor to a microcontroller board hosting the Arduino UNO R3 [53].

We purchased the ADNS-9800 optical sensor from a private independent supplier in a package including the sensor, the lens, and an accompanying application sensor board (see Figure 20).

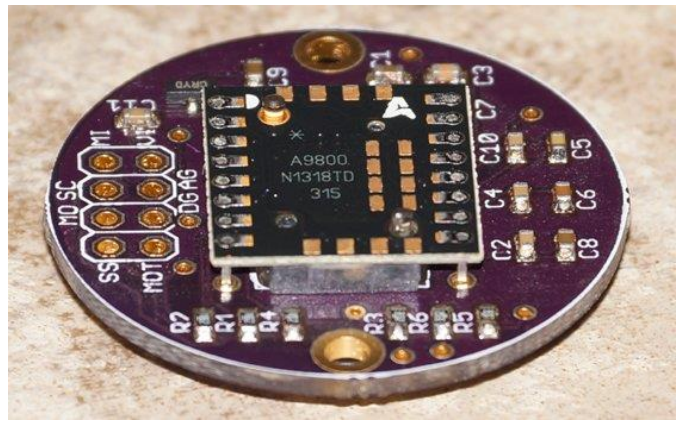

*Figure 20 - The optical sensor package purchased initially includes the purple application PCB, and the ADNS-9800 sensor on top.*

The optical sensor was then removed from the board using a ribbon cable, and placed underneath the fingertip as in the final prototype in Figure 19. The removal of the chip off the supporting

PCB led to power supply issues. We then provided the board with an extra 100pF capacitor, placed between the Arduino Uno and the application PCB in order to stabilize the voltage.

### 3.1.4   Computer Interfacing

The inertial measurement unit and optical sensor stream data to an Arduino UNO R3 board (see Figure 21). The Atmega16U2 microcontroller on Arduino then applies Kalman filtering to the data from the IMU, and synchronizes the orientation result with relative position data from the optical sensor. The fused data are then streamed to a computer, which is an HP ProBook 4530s running Ubuntu 12.04. An USB cable is used to connect the Arduino UNO to the computer for evaluation purposes. This wired connection later could be replaced by a wireless solution using a pair of XBee modules.

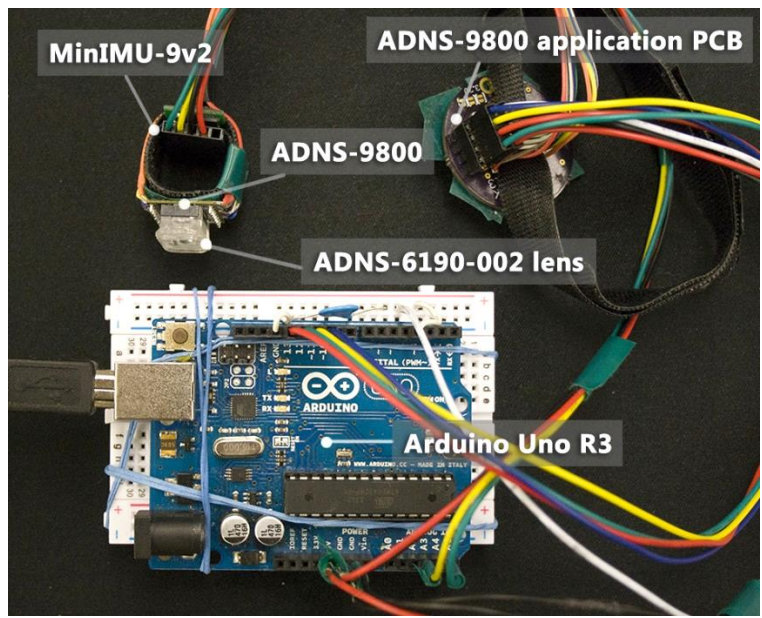

*Figure 21 - The entire setup of the hardware prototype.*

### 3.1.5 Lens Augmentation

To enable an additional feature of 3DTouch, we augmented a thin layer of elastic rubber of 2.0 mm height around curvature of the ADNS-6190-002 lens (see Figure 22). Without the rubber layer, the fixed distance from the lens to the surface is 2.4 mm. The rubber layer allows the distance from the lens to the touch surface to be adjustable from 2.4 to 4.4 mm by applying pressure. The distance from the sensor to the object surface ranges from 7.95 to 9.95 mm. This augmentation enables our gesture recognition engine to sense pressure as well.

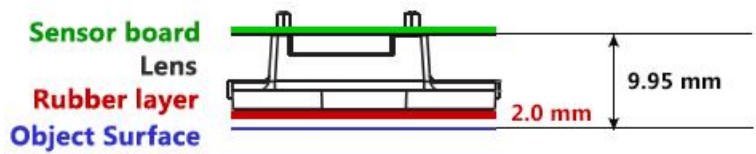

*Figure 22 - Diagram of the component layers of the optical sensor.*

## 3.2 Gesture Detection

For the device to be usable, we decided to implement the basic touch gestures of tap and double-tap. A novel *press gesture* is also proposed. This section explains the algorithms used to enable the tap, double-tap, and press gestures.

### 3.2.1 Sensing contact

To sense contact with a surface, Magic Finger relies on rapid changes in the pixel contrast level of the sensor image [40]. This approach requires continuous reading of the image pixels, and performing the calculation to derive the change in contrast level.

However, we took a simpler, yet effective approach by monitoring the surface quality (SQUAL) values reported directly by the ADNS-9800 sensor board. As described in the datasheet [41], the SQUAL register is a measure of the number of valid features visible by the sensor in the current frame. The following formula is used to find the total number of valid features:

$$Number\ of\ Features = SQUAL\ Register\ Value * 4$$

The maximum SQUAL register value is 169. Since small changes in the current frame can result in changes in SQUAL, variations in SQUAL when looking at a surface are expected. The graph below shows 800 sequentially acquired SQUAL values, while a sensor was moved slowly over white paper.

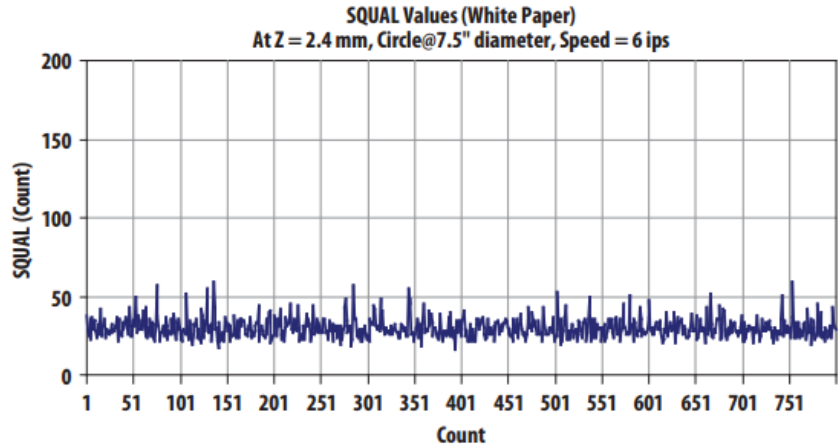

*Figure 23 - 800 sequentially acquired SQUAL values over a white paper. The range is approximately from 25 to 60. Image courtesy: ADNS-9800 datasheet*

SQUAL is nearly equal to zero if there is no surface below the sensor (Figure 24). SQUAL remains fairly high throughout the Z-height range which allows illumination of most pixels in the sensor.

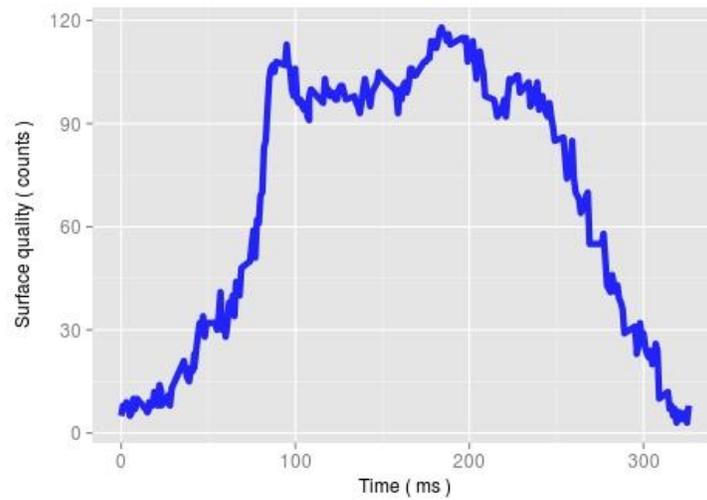

*Figure 24 - Surface quality values reported when users performing a tap gesture on the mousepad texture. SQUAL values stay above 90 when the optical lens is in contact with the surface.*

For each surface texture, a range of SQUAL values is provided. We used SQUAL in detecting tap gestures. However both approaches of using image contrast level, and SQUAL are still optical techniques to sense contact. Hence, the sensing accuracy is affected by such variables as environment lighting condition, surface texture, and the lift detection (Z-height) setting programmed to the optical sensor. Different surfaces will have different lift detection values with the same setting due to different surface characteristic [41].

### 3.2.2 Tap Gesture

The SQUAL, and x-y increments (e.g., DELTA_X, and DELTA_Y) values are used to measure tap gesture. A tap gesture is recognized when there is a rapid change, within 300 ms timespan, in SQUAL from 0 to 40, and in x/y movements between +/-5 units. These settings are specific values for mousepad texture only. When using a texture recognition engine based on Support-Vector Machines [40], it is possible to load the correct settings for corresponding textures.

### 3.2.3 Double-Tap Gesture

Similar to a double-click gesture, we needed to continuously monitor the tap gestures. If two tap gestures take place within a certain pre-defined time span, then a double-tap gesture is fired. Microsoft Windows 7 sets 500ms as the default time span for a double-click [54]. However, this should be an adjustable setting for users, and for the purposes of testing we set it to be 200-500ms.

On the other hand, for a double-tap gesture to be recognized, two subsequent taps need to take place at the same position. This is difficult to achieve with optical sensing because there is always noise when the sensor is lifted off the surface. After pilot testing 300 double-tap gestures, we defined the offset distance for two subsequent taps to be recognized as a double-tap to be +/- 15 for mousepad texture.

### 3.2.4  Press Gesture

The lift detection distance for ADNS-9800 ranges from 1-5 mm [41]. As the fixed height of the ADNS-6190-002 is 2.4 mm, the 2.0 mm thin layer of rubber allows the sensor to still recognize the surface within the 2.4 - 4.4 mm range. For the mousepad texture, an average SQUAL value of 40 corresponds to 2.4 mm lift-off distance under normal indoor light condition. We continuously monitor and detect a press gesture when the SQUAL values reach 40 or above.

This gesture reduces workload for the finger joint as users do not have to lift their finger off the surface. However, it is subject to many other environmental factors such as surface texture, and lighting condition. A mechanical push button may be a more reliable alternative.

## 3.3 3DTouch Interaction Techniques

This section describes how a single 3DTouch device, worn on a finger or thumb, can be used to perform 3D interaction techniques of selection, translation, and rotation. Interaction techniques utilizing more than one piece of 3DTouch are discussed in the Future Work section, and are not within the scope of this paper.

### 3.3.1 Selection

3DTouch is capable of sensing the absolute 3DOF orientation of the finger wearing the device. Hence, we propose to use the traditional Ray-Casting technique [6] to select an object in 3D space. In form factor 3 (section 3.1.1), 3DTouch is worn as a ring on the index finger. With ray casting, the user points the finger wearing 3DTouch at objects with a virtual ray that defines the direction of pointing (see Figure 25b). More than one object can be intersected by the ray, and only the one closest to the user should be selected. On a 2D plane such as the TV screen, the user can point up-down and left-right to move the 2D pointer around (see Figure 25a).

After a ray is pointed at an object, a tap gesture can be performed to make the selection command. For the selection technique, users can either wear the device as a ring (i.e., on the proximal phalanx), or as a thimble (i.e., on the fingertip).

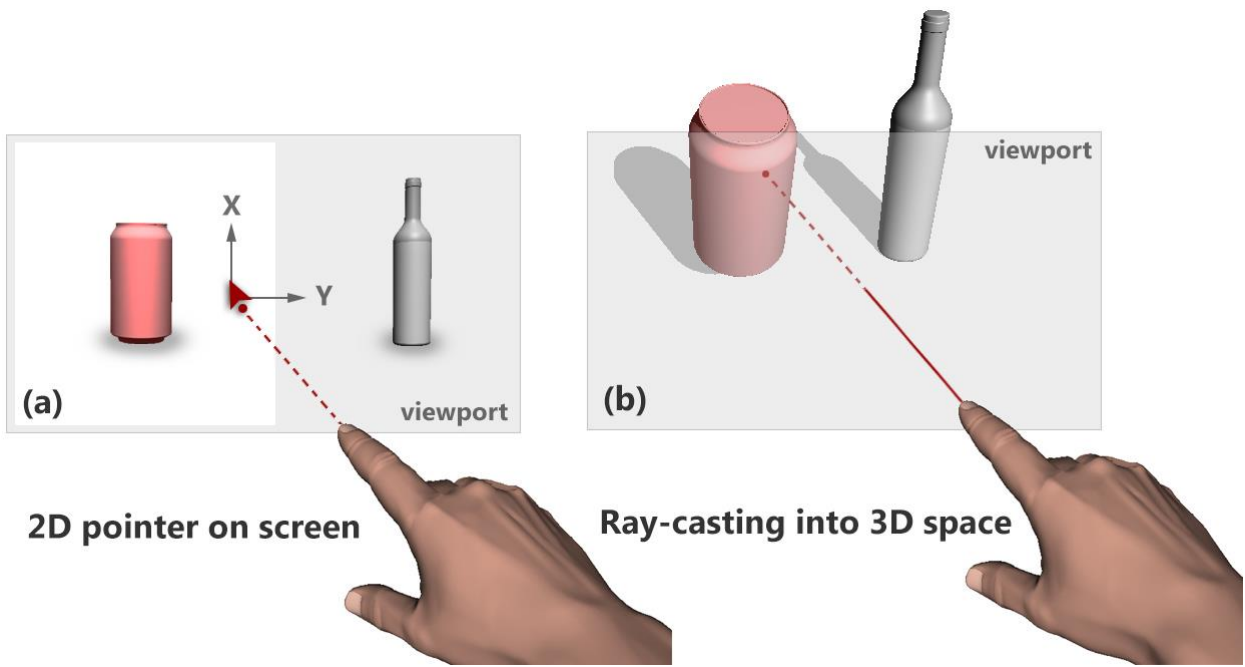

*Figure 25 - Selection technique using 3DTouch. (a) Moving the 2D pointer to the left half of the 2D plane to select the soda can. (b) Pointing at the soda can in 3D space to select it.*

## 3.3.2 Translation

With an optical sensor, 3DTouch is capable of drawing or translating an object on a 2D plane. However, this plane's orientation is not fixed, and is determined by the orientation of the finger. Figure 26 illustrates two examples of how the actual touch movements map to a 3D virtual environment (VE). This interaction technique can be applied to both object and screen translation.

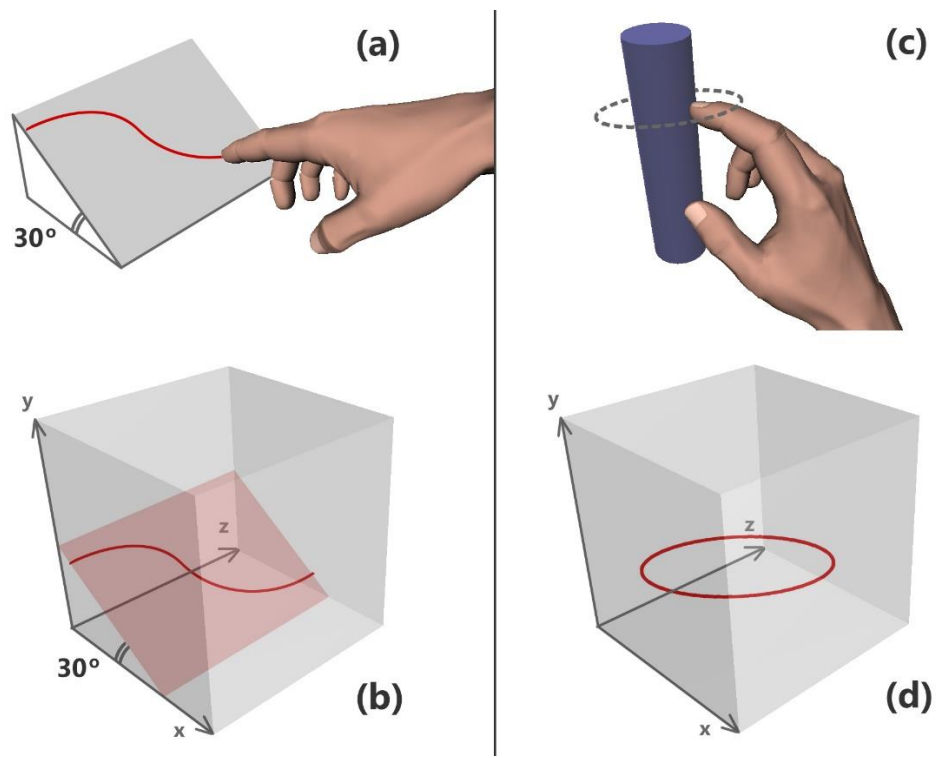

*Figure 26 – Translation technique using 3DTouch*

*(a) The 3DTouch user is drawing a curve (red) on a flat surface, which makes 30 degrees with the ground.*

*(b) In the 3D VE, a curve is generated on a 2D plane, which also makes 30 degrees with the XZ plane.*

*(c) The 3DTouch user is touching around the surface of a cylinder.*

*(d) In the 3D VE, a circle with diameter proportional to that of the cylinder is generated.*

With 3DTouch, users not only can interact with flat surfaces; however, they can interact with curved surfaces as well. This possibility, as described in Figure 26, opens up a whole new design space for novel 3D interaction techniques. Users can draw a curve line in the VE by simply performing touch interaction on the curve of a physical object. The touch interfaces that 3DTouch users interact with do not have to be flat, and so they can come in different shapes.

**Deriving the orientation of the 2D touch plane**

In form factor 1 (see Figure 15a), the IMU is perpendicular to the touch surface as illustrated in Figure 26a/c, we added 90 degrees to the *pitch* angle reported from the IMU to achieve the 2D plane orientation.

In form factor 2 (see Figure 15b), the IMU is parallel with the touch surface, the orientation of the 2D plane is equal to the orientation of the finger.

### 3.3.3 Orientation

Similar to translation, the user draws a vector on a surface to rotate a virtual object in focus. And the object will be rotated around the rotation axis, which is perpendicular to the drawn vector on the same 2D plane. The length of the vector drawn is proportional to the rotation angle. And the direction of the vector determines the rotation direction. Figure 27 illustrates an example of how the drawn vector is used to derive the rotation in a 3D VE.

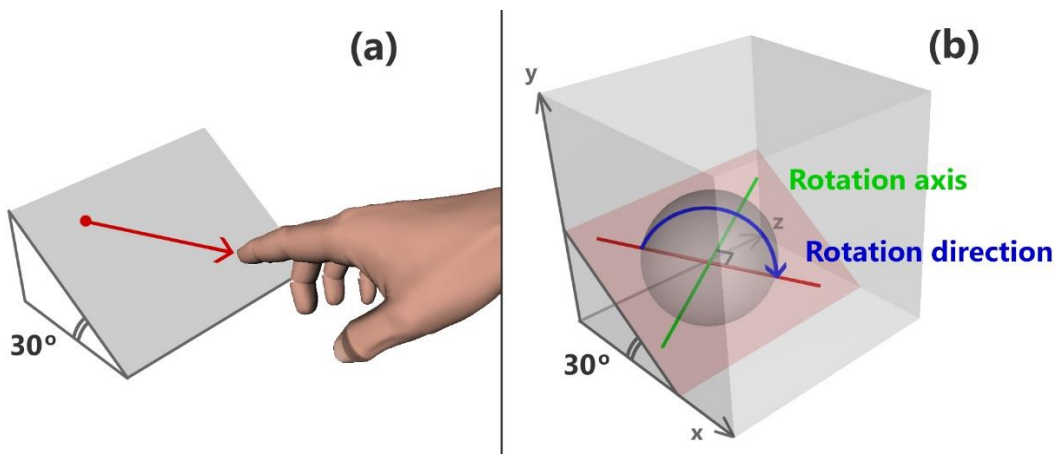

*Figure 27 - (a) The user is drawing a vector (red) on a flat surface, which makes 30 degrees with the ground. (b) In the 3D VE, the sphere is rotated around the rotation axis by an angle proportional to the vector length. The rotation axis is perpendicular to the vector on the 2D plane, which also makes 30 degrees with the XZ plane.*

## 3.4 Software Implementation

This section presents the software implementation that handles the signals sent from the sensors to the microcontroller, and from the microcontroller to a 3D visualization program on Linux (see Figure 28).

On the Arduino, we wrote an Arduino program, called a *sketch*, to synchronize the data from the sensors. A Kalman filter was applied to the orientation data from the inertial measurement unit. The synchronized data are streamed to our computer which runs a 3D application based on Virtual Reality User Interface (Vrui). We wrote the driver for Vrui to work with 3DTouch. Moreover, a Vrui *tool* was also written to perform 3D interaction techniques as described in section 3.3. Also, a Vrui application was written to demonstrate the usage the Vrui tool.

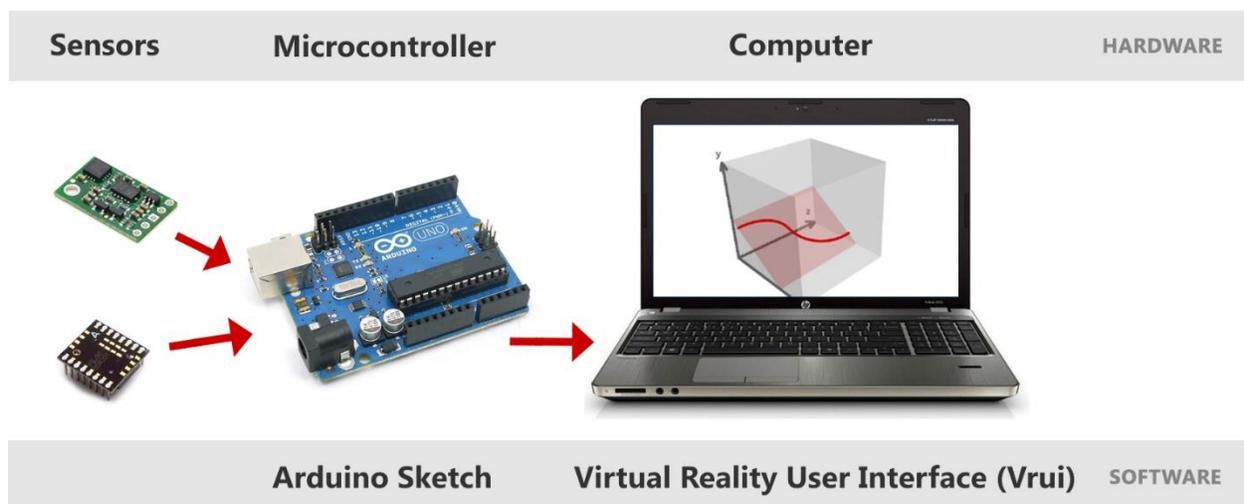

*Figure 28 - The hardware and software stacks of 3DTouch and its 3D demo application.*

### 3.4.1 Arduino sketch

A *sketch* is the name that Arduino uses for a program. It's the unit of code that is uploaded to and run on an Arduino board. On startup, our Arduino writes to the optical sensor with initial settings such as frame rate of 1500 fps, and resolution of 400 cpi.

At a frequency of 50 Hz, the sketch constantly reads data from the accelerometer, gyroscope, and compass of the IMU. The Kalman filter [55] is then applied to these values to help predict the 3D orientation stored in the Direction Cosine Matrix (DCM). Our Kalman filter implementation was based on the open-source code *sf9domahrs* by Doug Weibel and Jose Julio [56], and *ArduIMU v1.5* by Jordi Munoz and William Premerlani, Jose Julio and Doug Weibel [57]. The sketch then streams the orientation in quaternion notation [58] to the connected computer.

On the other hand, when there is motion detected by the optical sensor, the relative x and y movements will be sent to the computer together with the orientation above.

The sketch also detects tap and double-tap gestures by monitoring the surface quality and x/y movements as described in section 3.2.2.

We sent data from the Arduino to the computer in the following format:

| Orientation | | | | Position | | Gesture | |
|---|---|---|---|---|---|---|---|
| $q_x$ | $q_x$ | $q_x$ | $q_x$ | x | y | Tap | Double-tap |
| *float* | *float* | *float* | *float* | *int* | *int* | *char* | *char* |

*Table 1 - Format of output generated by Arduino sketch*

For the gesture representation, we used "X, O" to represent a tap, and "X, X" to represent a double-tap. A sample output from Arduino would be *"0.2, 0.4, 0.1, 0.4, -2, -4, X, O"*.

### 3.4.2 Virtual Reality User Interface application

On the computer side, we chose Virtual Reality User Interface (Vrui) as the 3D application framework to demonstrate our input device.

Vrui is a C++ development toolkit for highly interactive and high-performance VR applications, aimed at producing completely environment-independent software [59]. The reasons we chose Vrui are because it is free, and open-source. This enables us to write drivers and extensions for Vrui to work with our input device. On the other hand, Vrui applications run effectively on widely different VR environments, ranging from desktop systems with only a keyboard and a mouse to fully-immersive multi-screen systems. This allows us to further demonstrate the usability of 3DTouch in different VR environments beside desktops.

We used the latest version of Vrui 3.1 which can be downloaded at [60]. We wrote the driver for Vrui to work with 3DTouch. Also, a Vrui application was written to visualize the 3D position and orientation of 3DTouch for testing, and to demonstrate the interaction techniques of the input device as described in section 3.3.

### 3.4.3 How to fuse and derive 3D position from the sensory data?

The Optical Sensor: gives us a pair of 2D position increments (x, y) at a time *t*. These values, x, and y measure the distance in counts that the sensor has traveled over a delta time $t_2 - t_1$ on a 2D plane.

The Inertial Measurement Unit: gives us the absolute 3D orientation (relative to the ground that the user is standing on) of the 2D plane on which the optical sensor moves on. This 2D plane is the touch surface that the sensor is in contact with.

At every point in time, we can determine the relative 3D position of the optical sensor on a certain 2D surface. The figure below visually demonstrates the concept behind the data fusion described above:

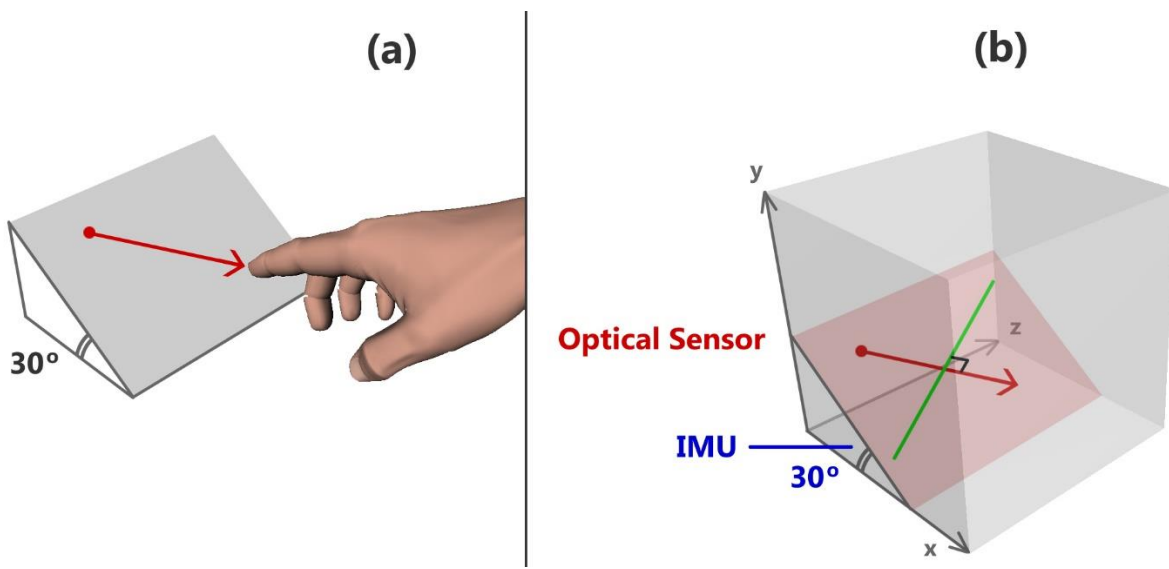

*Figure 29 - The IMU determines the orientation of the 2D plane on which the optical sensor moves on*

*The optical sensor determines the direction and magnitude of movement on the 2D plane (touch surface).*

# 4  Evaluation

We conducted an experiment to evaluate the 3D tracking accuracy of our device with respect to the ground truth across multiple surfaces. We compared the 3D position and 3D orientation reported by 3DTouch, against the data obtained using NaturalPoint OptiTrack motion tracking system [61]. In this experiment, we assumed the data obtained from the OptiTrack system to be the ground truth. The mean error of OptiTrack ranges from 0.1mm to 0.8mm according to our observations during the experiments.

## 4.1  Setup

3DTouch and OptiTrack both streamed their data via wired connections to a Linux machine with a dual-core 2.1GHz CPU with 4GB of RAM. On this machine, a program written in C++ synchronized and logged the samples at 50Hz. The device was configured in form factor 1, and worn by the author on the index finger.

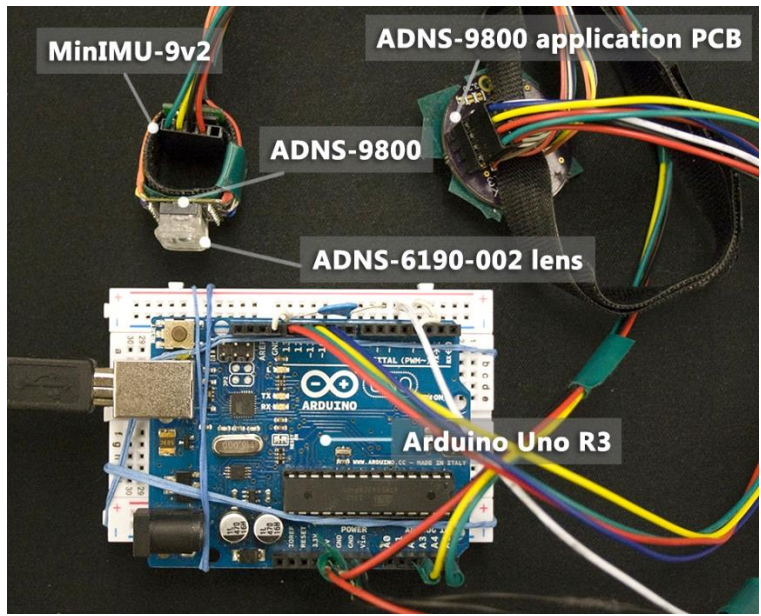

*Figure 30 - The setup of 3DTouch including the Arduino UNO R3, an optical sensor, and IMU, and a sensor application board (purple).*

To capture the movements of 3DTouch, we setup the tracking volume using 12 Flex-13 cameras sampling at 120 Hz (see Figure 31). The cameras are positioned circularly around the testing desk where the researcher sits to perform the experiment.

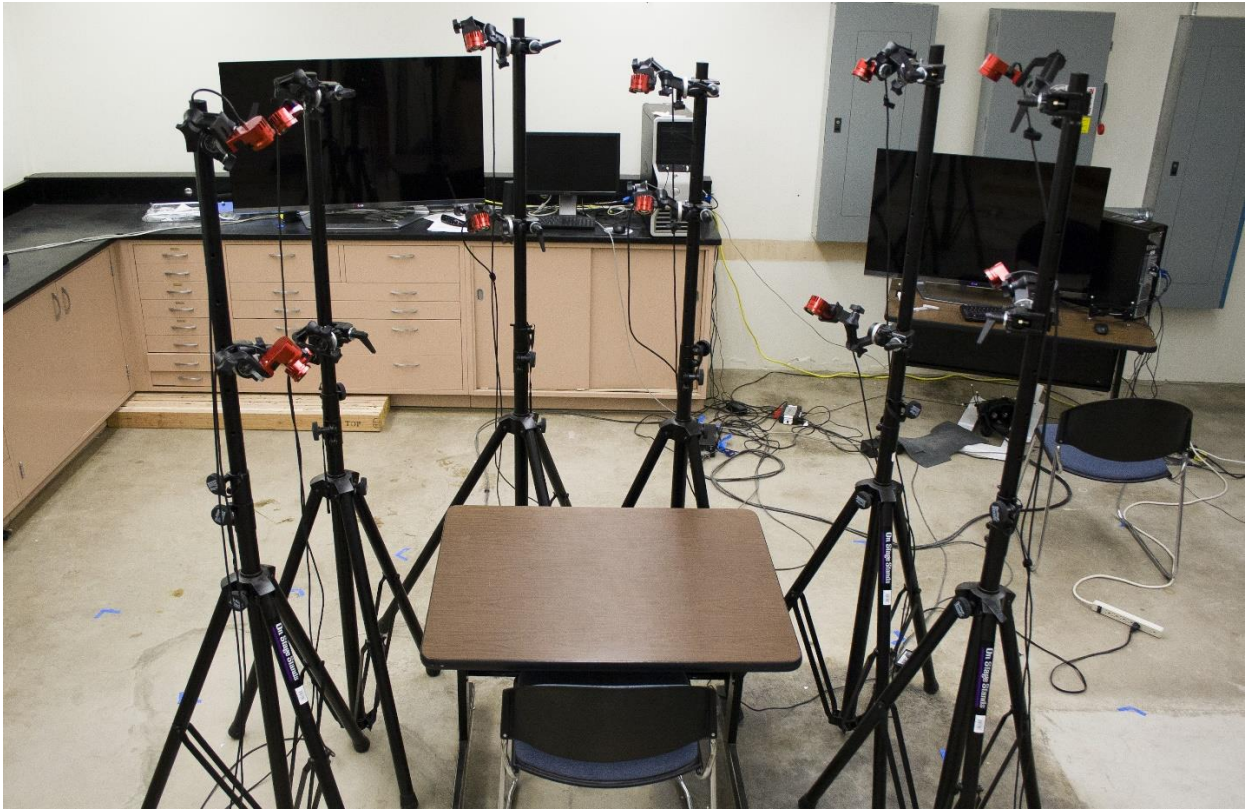

*Figure 31 - 12 Flex13 cameras (red) are positioned around the desk on the experimental setup.*

On the table, the researcher performs subtle touch movements as described in section 4.2. The movements are recorded by 3DTouch and the OptiTrack system and compared against each other.

In order for the OptiTrack to track the movements of our device, a rigid body, composed of three reflective markers, was mounted on top of 3DTouch as in Figure 32. The OptiTrack shoots out an invisible infrared ray into the tracking volume, and the reflective markers reflect back to the cameras. For more details of how the optical tracking system works, see section 2.1.3.

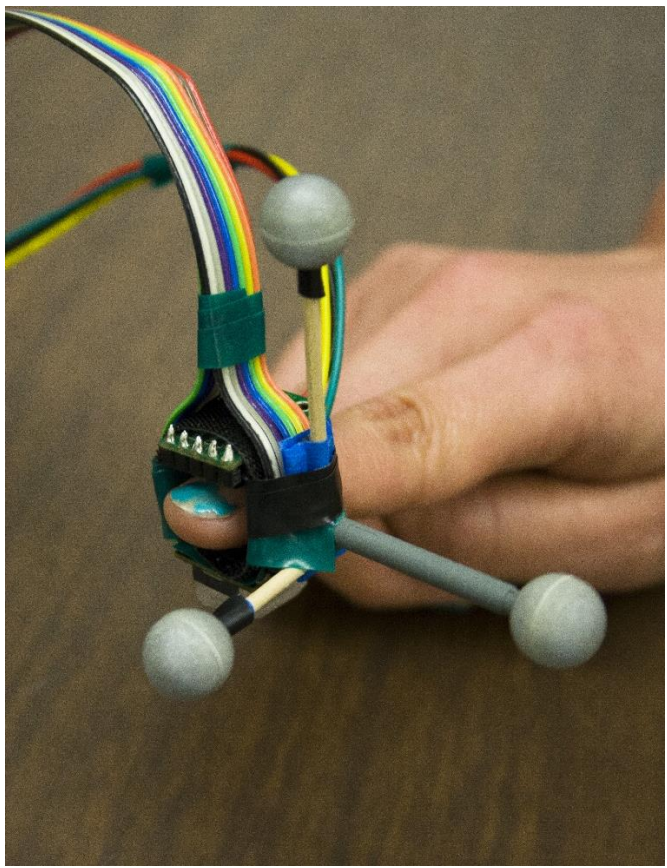

*Figure 32 - 3DTouch with reflective markers on for optical tracking*

## 4.2 Experimental Design

Since the surface texture is a factor affecting the optical sensing accuracy, we tested the device across 3 textures: mousepad, wooden desk, and jeans. These are three of the environmental textures used as contextual input for Magic Finger [40]. And these three textures are commonly found in an office environment matching our target use case of 3DTouch (see Figure 33).

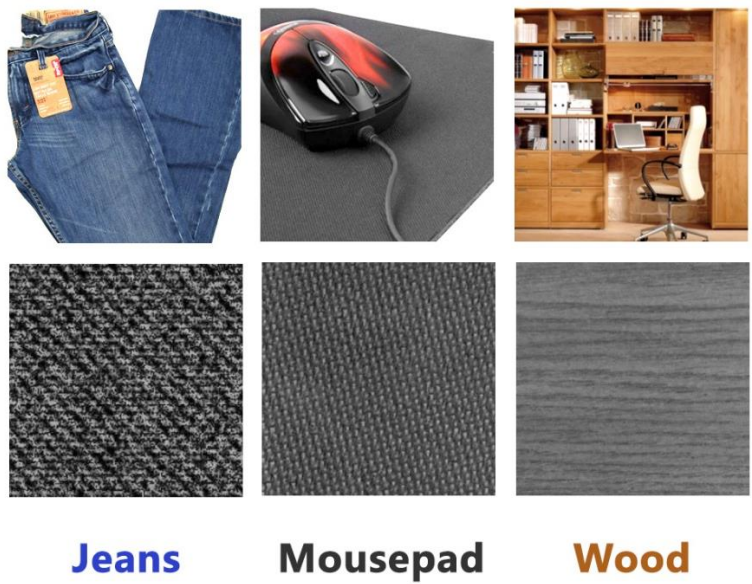

*Figure 33 - Three textures of Jeans, Mousepad and Wood that we test 3DTouch on. The lower row shows the images of these textures captured by the optical sensor.*

For each texture, we designed 4 different target area sizes: 12 x 12mm, 21 x 21mm, 42 x 42mm, and 84 x 84mm (see Figure 34). We chose 12 x 12mm as the smallest size because that is the smallest touch area usable by a previous work [36]. The largest area is designed according to the average human palm size [62], which is the touch area for the target mobile applications of 3DTouch.

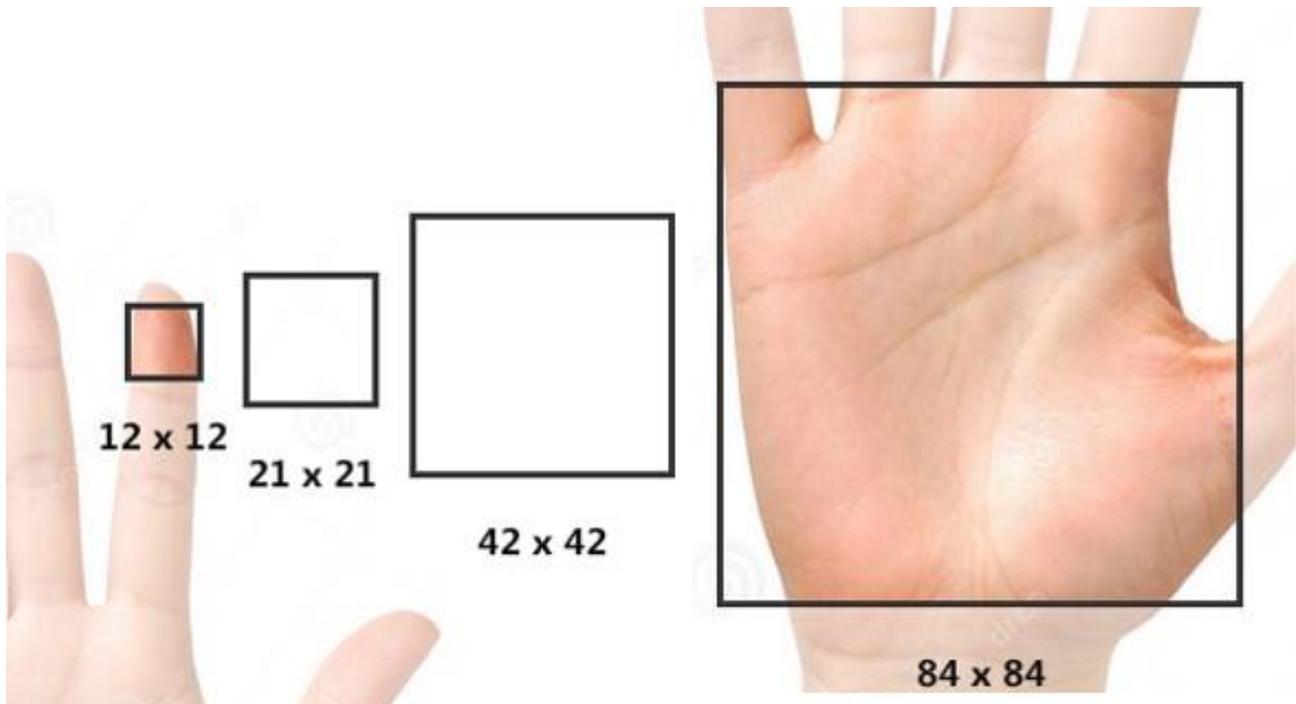

*Figure 34 - Actual target sizes (in millimeter square). Ratio 1:1*

For each target size, we performed the drawing 6 basic shapes: horizontal line, vertical line, diagonal line, triangle, square, and circle (see Figure 35). These basic shapes are the building blocks for users to perform 3D interaction techniques and 2D gestures. The horizontal lines test the tracking of X-axis movements of the optical sensor. The vertical lines test the tracking of Y-axis movements. The diagonal lines in the shapes test both X, Y tracking simultaneously.

In total, the experiment design was 3 x 4 x 6 (Texture x Size x Shape) with five repetitions for each cell to minimize the human error factor. For each drawing task, the touch surface is tilted at a random angle within 0 to 90 degrees from the ground. This random plane orientation setup tests the inertial measurement unit.

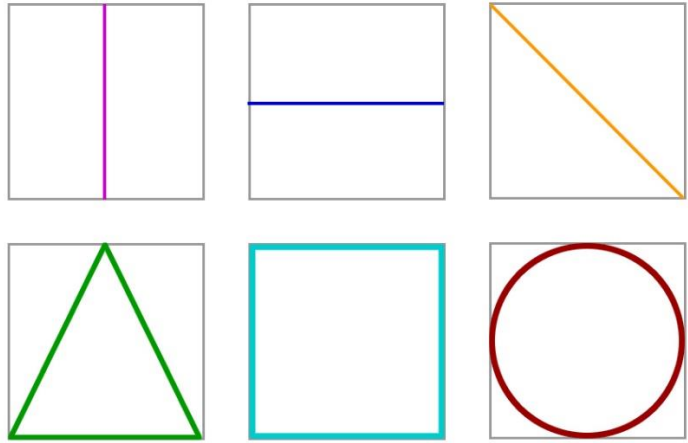

*Figure 35 - Six different target shapes for each target size.*

## 4.3 Results

There were 72,000 data points collected in total. We measured the Euclidean error in 3D position and 3D orientation of the directional vector for every data point reported by 3DTouch and OptiTrack.

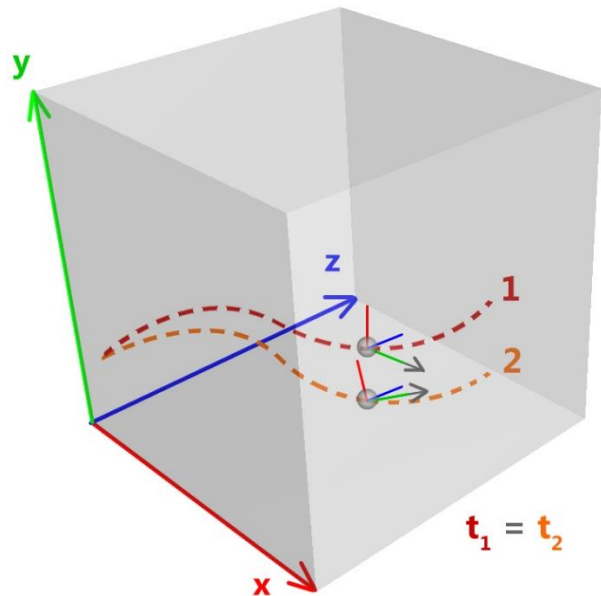

*Figure 36 - For every point in time, we compare the position and orientation of a point by 3DTouch and a point by OptiTrack. The directional vectors are colored grey.*

Because 3DTouch reports relative position, the paths generated by 3DTouch and OptiTrack are aligned to start at the same position (see Figure 36). For every point in time, we compare the Euclidean difference between 3D positions of points generated by two systems. To compare the orientation of these points, the angles between two directional vectors which are on their respective y axes, are calculated. Hence, for every time *t*, we have a position difference measured in millimeters, and an orientation difference measured in degrees.

### 4.3.1 Overall accuracy results

The overall mean position error is 1.10 mm ($\sigma = 0.87$), and the orientation error is 2.33 degrees ($\sigma = 2.58$). The position error is a high overall error given the small target area sizes. As a relative reference, optical mice with similar resolution of 400 cpi, and frame rate of 1500 fps used in mobile robot odometry measurement had the maximum error below 0.8 mm in a 50 mm range [63].

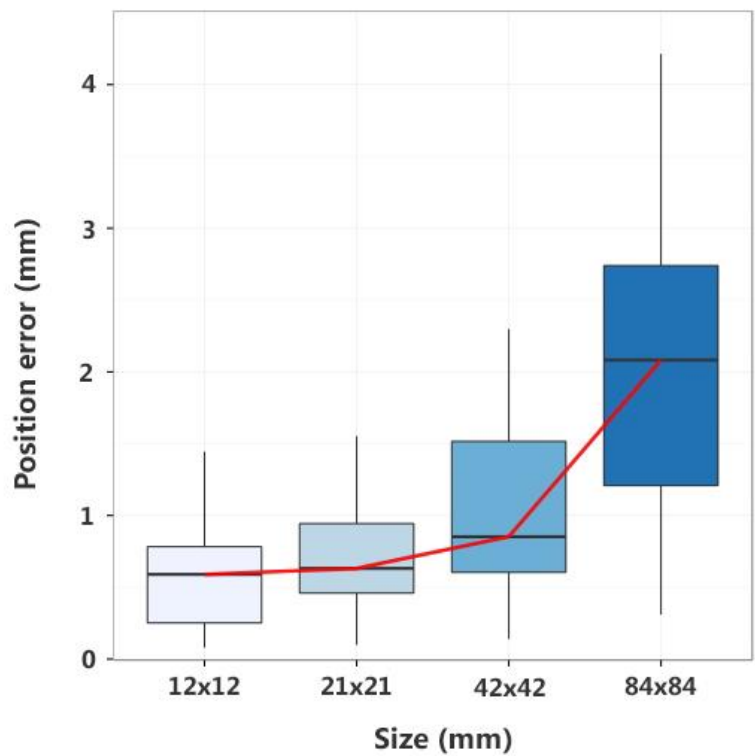

*Figure 37 - Mean position error increases as the target size increases*

Since the optical sensor reports relative position, it is subjected to drift over a long travel distance. The mean position error increases as the target size increases (see Figure 37). A similar trend is found with the orientation error (see Figure 38). This phenomenon can be explained due to the relative positioning technique that 3DTouch uses.

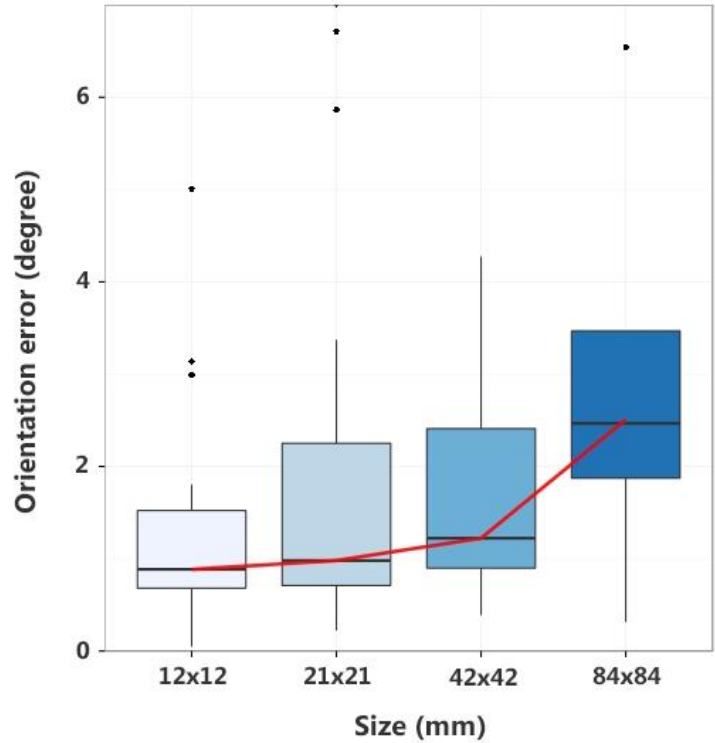

*Figure 38 - Mean orientation error does not always increase as the target size increases.*

The three textures tested all have a high surface quality between 30 and 50, and did not show significant difference in position error (see Figure 39). Table 4 shows the mean errors across three tested textures. In the future work, we would like to test 3DTouch on a larger number of textures. We ran Analysis of Variance (ANOVA) on the errors across three different texture groups, and we did not find significant difference with $p$-value = 0.116 and $F$ = 2.227 (see Table 2).

|  | DF | Sum square | Mean square | F value | P-value |
|---|---|---|---|---|---|
| **Group** | 2 | 3.331 | 1.66543 | 2.227 | 0.1156 |
| **Residuals** | 69 | 51.602 | 0.74785 |  |  |

*Table 2 - ANOVA result table of testing error across three different textures.*

Table 3 summarizes the errors of position and orientation across target sizes:

|  | **12x12 mm** | **21x21 mm** | **42x42 mm** | **84x84mm** |
|---|---|---|---|---|
| **Position error (mm)** | 0.61 | 0.70 | 1.02 | 2.05 |
| **Orientation (degree)** | 1.36 | 1.98 | 1.73 | 4.23 |

*Table 3 - Mean error across target sizes*

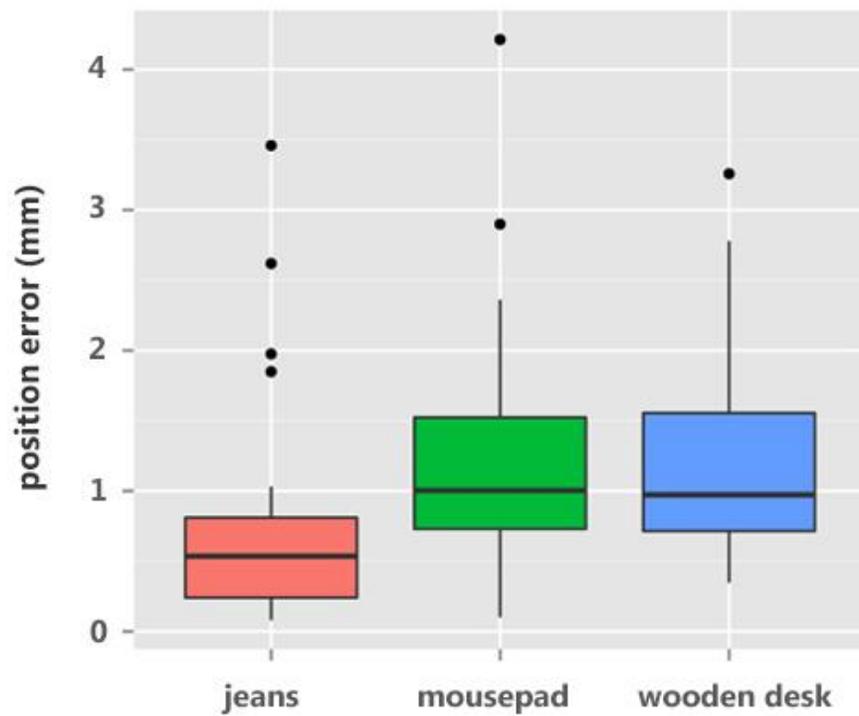

*Figure 39 - Mean position errors on jeans, mousepad and wooden desk textures are not significantly different.*

|  | **Mousepad** | **Jeans** | **Wooden Desk** |
|---|---|---|---|
| **Position error (mm)** | 1.27 | 0.79 | 1.23 |

| Orientation (degree) | 1.13 | 4.56 | 1.28 |

Table 4 - Mean errors across three different textures

### 4.3.2 Problem on acceleration

We also realized a number of data clouds with high error generated by 3DTouch have largely variable distances from point to point. This suggests our position error is also partly due to the inherent acceleration error (i.e., up to 30g) in the ADNS-9800 sensor.

The following pictures show some of the results with mean errors above 15 mm.

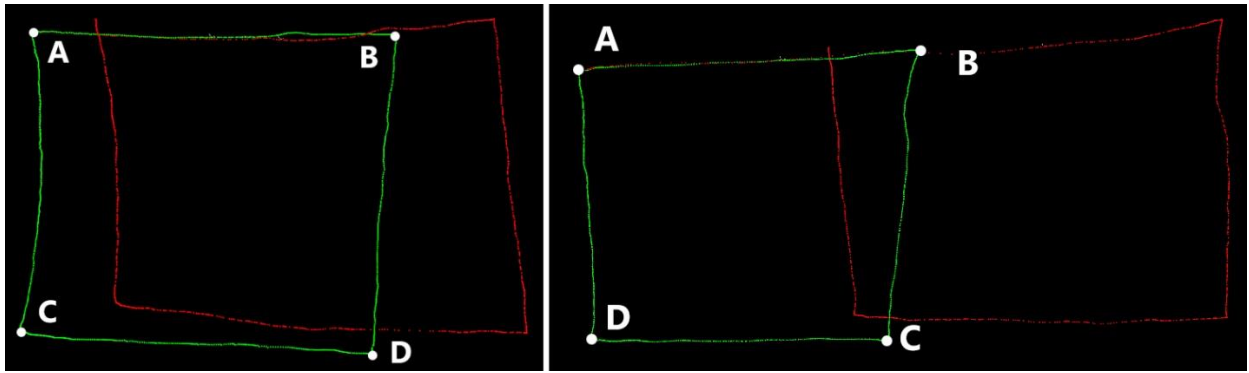

*Figure 40 - Example of results of 84x84mm square tests with high mean errors above 15mm.*

**Green** *dots are 3D points reported by OptiTrack, which is assumed to be our actual finger movements.*

**Red** *dots are 3D points reported by 3DTouch.*

*The squares were drawn in the direction from A → B → C → D.*

Figure 40 shows two visual results of 84x84mm square tests with high mean errors above 15 mm. The green dots are 3D points reported by OptiTrack, which is assumed to report our actual finger movements. The red dots are 3D points reported by 3DTouch. We can see from Figure 40 that although the A points of green and red rectangles are aligned at the same position, the segment AB drawn by 3DTouch is largely longer than its own CD segment. Since the movements were conducted on the same texture under a consistent lighting condition, the results suggest that the inherent acceleration problem in the optical sensor caused such false results. The readings of relative position reported by the optical sensor were affected by the movement speed up to 30g acceleration [41].

This suggests that a more reliable optical sensor such as ADNS-2030 with 0.5g acceleration may significantly improve the performance. Hence, the results presented in this document should be the baseline accuracy.

## 4.4 Further evaluation on precision

We performed a further precision test to verify the acceleration error discovered in section 4.3.2. We would like to prove that there is a significant proportional correlation between the movement speed and the optical sensor readings (in counts). Acceleration is a big problem that affects both accuracy and precision of the device as the outputs are not consistent across different speeds. We would like to know:

- At what resolution this acceleration problem happens.
- At which speeds this acceleration problem happens.
- At what magnitude this acceleration problem happens.

Knowing the answers to the above questions would help us decide to either fix the problem in the microcontroller level or replacing the optical sensor.

### 4.4.1 Experimental Design

We disassembled our device and mounted the optical sensor underneath a four-wheeled toy vehicle. This vehicle allows the optical sensor to be at a fixed vertical distance of 5mm from the mousepad texture surface. The toy vehicle is restricted to only move forward and backward 1 inch. This restricted region allows us to move the vehicle at different speeds over the 1 inch distance and measure counts reported by the optical sensor (see Figure 41). In a reliable optical sensor, moving at different speeds over the same distance would result in similar counts.

We performed two conditions, one at 400 cpi, and one at 800 cpi resolution in order to verify the acceleration problem happens at 400 cpi and higher. For each condition, the vehicle was moved 500 times forward and backward resulting in 500 trial sample readings. An Arduino program was written to record the speed of movement over 1 inch, and the counts reported by the optical sensor.

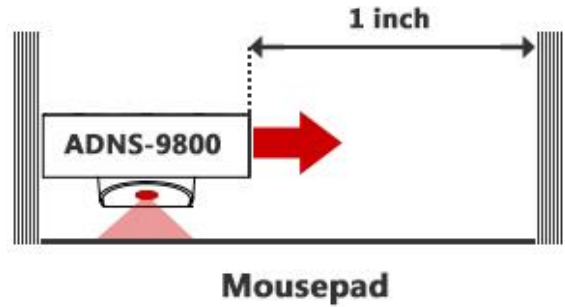

*Figure 41 - The ADNS-9800 optical sensor is moved horizontally back and forth over a fixed distance of 1 inch. The mousepad is used as the testing texture.*

### 4.4.2  Results on precision test

We collected over 1000 samples of counts and movement speeds reported in ips (inch per second). Figure 42 shows the counts reported by 3DTouch over 7 speeds ranges from 0 to 7. We can see the trend of acceleration from 0 – 3 ips. Beyond 3 ips, the counts start going back down slowly. We did not evaluate faster movements at above 7 ips because they are likely to result in instability of movements. Hence, they not practical gestures in applications that require accuracy.

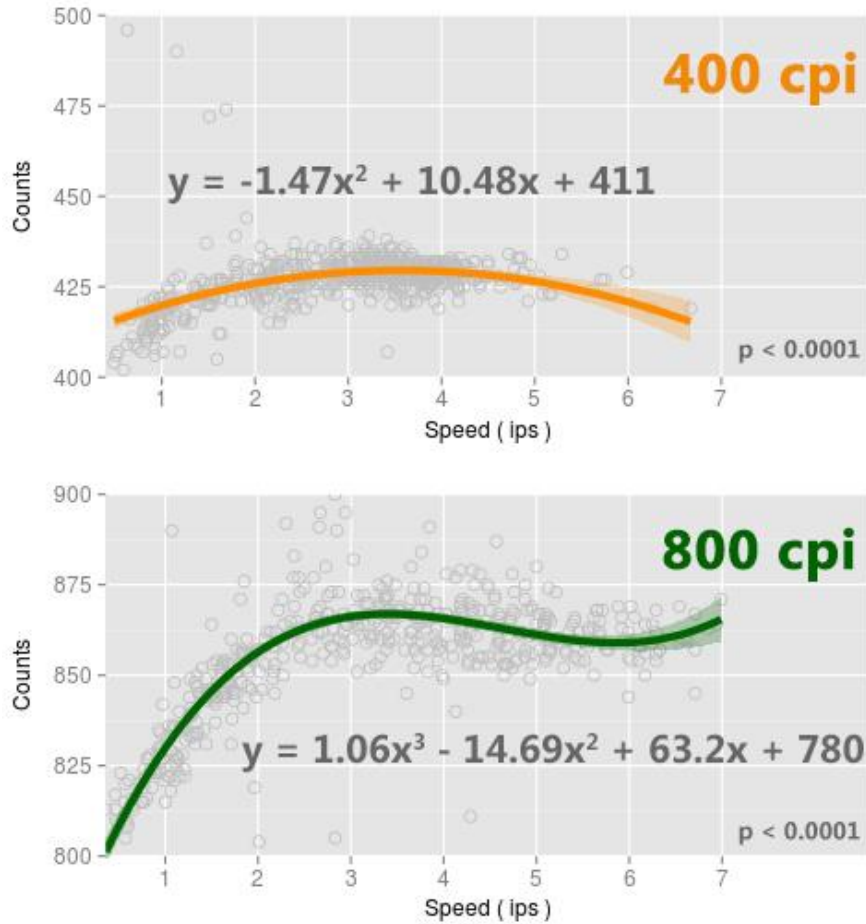

*Figure 42 - A total of 1000 samples obtained from the optical sensor. 500 were obtained at 800 cpi trial, and the other 500 were obtained from the 400 cpi trial.*

We fitted the data points generated at two conditions to non-linear regression models. The counts at 400 cpi were fitted successfully to a quadratic function ($y = -1.47x^2 + 10.48x + 411$) with significance ($p < 0.0001$). And at 800 cpi, the counts were fitted successfully to a cubic function ($y = 1.06x^3 - 14.69x^2 + 63.2x + 780$) with significance as well ($p < 0.0001$). Knowing these two functions may help us fix the problem by applying a filter to the data from the optical sensor (see Table 5).

|  | Counts | Speed | Speed$^2$ | Speed$^3$ | R-squared | p-value |
|---|---|---|---|---|---|---|
| **400 cpi** | 410.8021 | 10.4840 | -1.4685 |  | 0.1613 | 2.2e-16 |
| **800 cpi** | 780.3591 | 63.1833 | -14.6930 | 1.0582 | 0.6444 | Less than 2.2e-16 |
|  | *\* p-values for all the coefficients are less than 0.0001.* |  |  |  |  |  |

*Table 5 – Non-linear functions successfully fitted to the counts at 400cpi and 800 cpi.*

To prove that the acceleration magnitude is significant, we ran regression analysis between speed and counts on the samples within 0 and 3 ips. This is the range that has the highest acceleration. At the resolution of 400 cpi, the counts accelerate in the range [415, 430]. And at the resolution of 800 cpi, the acceleration is even higher, from 815 to 870 counts. We found the significance p < 0.0001 for both 400 cpi and 800 cpi trials. The correlation slope for the higher resolution of 800 cpi shows to be 24.22, steeper than 5.64 of 400 cpi.

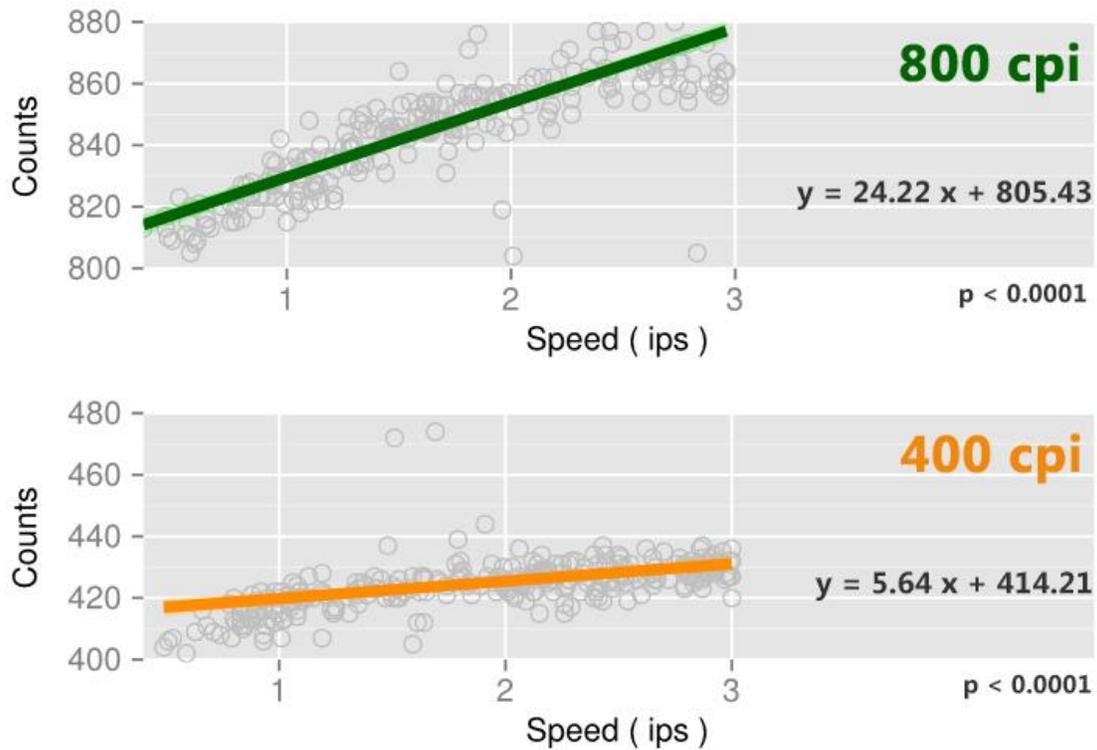

*Figure 43 - The linear correlation between speed and counts was proved to be significant within 0-3 ips.*

The table below shows the detailed regression analysis results from our R program:

|  | **Intercept (counts)** | **Slope (speed)** | **R-squared** | **p-value** |
|---|---|---|---|---|
| **400 cpi** | 414.2103 | 5.6379 | 0.1174 | 3.961e-09 |
| **800 cpi** | 805.430 | 24.215 | 0.6444 | Less than 2.2e-16 |
|  | *\* p-values for all the coefficients are less than 0.0001.* |  |  |  |

*Table 6 - Regression analysis between speed and counts from R program.*

### 4.4.3 Conclusion on precision test

The precision test presented in the previous section shows that there is a significant acceleration problem in ADNS-9800 sensor. Especially when moving at a speed within 0-3 ips on a mousepad texture, the optical sensor shows a max of 15 counts variation at 400 cpi, and 55 counts variation at 800 cpi. The overall mean accuracy found in the second experiment is 1.65mm ($\sigma = 0.55$) at 400 cpi, and 1.71mm ($\sigma = 0.57$) at 800 cpi.

This is a factor affecting both the accuracy and precision of our device when being used at various speeds.

# 5  Future work

The ADNS-9800 optical sensor proves to have an inherent hardware acceleration problem, which cannot be completely removed at the microcontroller unit level. For further accuracy and precision evaluation, we would like to replace this sensor by another more reliable sensor with less acceleration.

In this prototype, only one single 3DTouch device is used to wear on a finger. However, to enable multi-touch capability, it is possible to use more than one device on multiple fingers such as two index fingers, or index finger and thumb.

With 3DTouch, we explored implementing the gestures using the characteristics of optical sensors. This method may be novel relative to approaches using mechanical buttons, or capacitive touch technology. However, a further evaluation on the gesture recognition is necessary to investigate its performance in outdoor environments and across different textures.

Our accuracy evaluation showed that 3DTouch is capable of performing 3D translation with the mean error of 1.10 mm and 2.33 degrees within 84x84 mm target mobile touch area. However, a user study will be further conducted to measure usability feedback, especially fatigue and comfort level. We especially, would like to compare the fatigue and comfort level of 3DTouch against existing 3D input devices.

Also, we would like to further evaluate the performance of 3DTouch against the existing 3D input devices such as Wiimote, and mobile touch devices across different 3D virtual environment platforms (e.g., desktop, home theater, or CAVE). With touch interaction concept employed, 3DTouch can also be used with mobile touch devices with added value of finger orientation. This may be useful for certain applications that take finger orientation into account.

Beyond 3D applications, 3DTouch can serve as a touch and gestural input device in everyday life. The inertial measurement unit allows the finger to serve as pointing device. Moreover, the optical sensor adds more functionality with tap, double-tap and even custom gestures if incorporated with a machine learning algorithm such as Support Vector Machines.

# 6 Conclusion

In this paper, we presented a novel 3D wearable input device using a combination of a laser optical sensor, and a 9-DOF inertial measurement unit. 3DTouch enables users to use their fingers or thumb as a 3D input device with the capability of performing 3D selection, translation, and rotation. This device is designed to work universally across different platforms such as a desktop, CAVE, mobile touch devices and many more. Our evaluation, shows the device's overall average tracking accuracy of 1.10 mm and 2.33 degrees within four target sizes from 12x12 mm to 84x84 mm, and across three different textures of jeans, mousepad, and wooden desk. Both the first and second experiments showed the overall average tracking accuracy of below 2.0mm.

3DTouch is self-contained, and can be universally used on various 3D platforms. We described how we derive the 3D position using the 2D relative position from the optical sensor and 3D absolute orientation from the IMU. With 3DTouch, we attempted to bring 3D interaction and applications a step closer to users in everyday life.